\documentclass[apj]{emulateapj}
\usepackage{apjfonts}
\usepackage{natbib}
 
\usepackage{epsfig}
\usepackage{amssymb}
\usepackage{natbib}
\usepackage{aas_macros}
\usepackage{comment}
\usepackage{url}

\slugcomment{Accepted for publication in ApJ}

\shorttitle{Bulge growth CANDELS}
\shortauthors{Huertas-Company et al.}

\begin{document}
 
\voffset=-0.60in






\title{The morphologies of massive galaxies from $z\sim3$ - Witnessing the 2 channels of bulge growth. }


\author{\textsc{M. Huertas-Company\altaffilmark{1}, P. G. P\'{e}rez-Gonz\'{a}lez\altaffilmark{2}, S. Mei\altaffilmark{1}, F. Shankar\altaffilmark{3}, M. Bernardi\altaffilmark{4}, E. Daddi\altaffilmark{5},  G. Barro\altaffilmark{6}, G. Cabrera-Vives\altaffilmark{7,8}, A. Cattaneo\altaffilmark{1},P. Dimauro\altaffilmark{1}, R. Gravet\altaffilmark{1}}} 

\affil{{\scriptsize $^1$ GEPI, Observatoire de Paris, CNRS, Universit\'e Paris  Diderot, 61, Avenue de l'Observatoire 75014, Paris  France}}
\affil{{\scriptsize $^2$ Departamento de Astrof\'isica, Facultad de CC. F\'isicas, Universidad Complutense de Madrid, E-28040 Madrid, Spain ; Associate Astronomer at Steward Observatory, The University of Arizona}}
\affil{{\scriptsize $^3$ School of Physics and Astronomy, University of Southampton, Southampton SO17 1BJ, UK}}
\affil{{\scriptsize $^4$ Department of Physics and Astronomy, University of Pennsylvania, Philadelphia, PA 19104, USA}}
\affil{{\scriptsize $^5$ Laboratoire AIM, CEA/DSM-CNRS-Université Paris Diderot, Irfu/Service d'Astrophysique, CEA Saclay, Orme des Merisiers, 91191 Gif-sur-Yvette Cedex, France}}
\affil{{\scriptsize $^6$ University of California, Santa Cruz, 1156 High Street, Santa Cruz, CA 95064, USA}}
\affil{{\scriptsize $^7$ AURA Observatory in Chile, La Serena, Chile}
\affil{{\scriptsize $^8$ Department of Computer Science and Center for Mathematical Modeling, University of Chile, Santiago, Chile}}
}




\begin{abstract}
We quantify the morphological evolution of $z\sim0$ massive galaxies ($M_*/M_\odot\sim10^{11.2\pm0.3}$) from $z\sim3$ in the 5 CANDELS fields. The progenitors are selected using abundance matching techniques to account for the mass growth. The morphologies of massive galaxies strongly evolve from $z\sim3$. At $z<1$, the population well matches the massive end of the Hubble sequence, with $30\%$ of pure spheroids, $50\%$ of galaxies with equally dominant disk and bulge components and $20\%$ of disks. At $z\sim2-3$ however, there is a majority of irregular systems ($\sim60-70\%$) with still $30\%$ of pure spheroids.

We then analyze the stellar populations, SFRs, gas fractions and structural properties for the different morphologies independently. Our results suggest two distinct channels for the growth of bulges in massive galaxies. 

Around $\sim30-40\%$ were already bulges at $z\sim2.5$, with low average SFRs and gas-fractions ($10-15\%$), high Sersic indices ($n>3-4$) and small effective radii ($R_e\sim1$ kpc) pointing towards an even earlier formation through gas-rich mergers or violent disk instabilities. Between $z\sim 2.5$ and $z\sim0$, they rapidly increase their size by a factor of $\sim4-5$, become all passive and slightly increase their Sersic indices ($n\sim5$) but their global morphology remains unaltered. The structural evolution is independent of the gas fractions, suggesting that it is driven by ex-situ events.  

The remaining $60\%$ experience a gradual morphological transformation, from clumpy disks to more regular bulge+disks systems, essentially happening at $z>1$. It results in the growth of a significant bulge component ($n\sim3$) for $2/3$ of the systems possibly through the migration of clumps while the remaining $1/3$ keeps a rather small bulge ($n\sim1.5-2$). The transition phase between disturbed and relaxed systems and the emergence of the bulge is correlated with a decrease of the star formation activity and the gas fractions suggesting a \emph{morphological quenching} process as a plausible mechanism for the formation of these bulges (although the eventual impact of major mergers and a growing black hole in the bulge should also be considered). The growth of the effective radii scales roughly with $H(z)^{-1}$ and it is therefore consistent with the expected growth of disks in galaxy haloes. 

\end{abstract}



\section{Introduction}

In the local Universe massive galaxies are characterized by
having a dominant early-type, bulge-dominated morphology as well
as old stellar populations. They are also confined to tight scaling relations, such as the mass-size relation (e.g. \citealp{2003MNRAS.343..978S, 2014MNRAS.443..874B}) and the fundamental plane. Understanding the formation and subsequent mass assembly of such systems is however still debated in present-day cosmology and it is a key milestone towards reaching a complete view of structure formation and the interplay between baryons and their dark-matter hosts. In particular, the actual role played by mergers as compared to in-situ processes in shaping spheroids is still unclear, and state-of-the-art semi-analytic models of galaxy formation offer sometimes quite different views (e.g., \citealp{2009MNRAS.397.1254G, 2011ApJ...742...24L})

Following the scaling relations of these massive objects across cosmic time is a natural way to better understand how the relations actually emerged and the role played at different cosmic epochs by the different physical mechanisms. As a matter of fact, many works in the last ten years have focused their attention on the evolution of the mass-size relation for a selection of massive galaxies ($log(M_*/M_\odot)>10.5$) finding an apparent increase of the zero point of the relation by a factor of a few from $z\sim3$ (e.g~\citealp{2005ApJ...626..680D,2006ApJ...650...18T,2008ApJ...687L..61B,2008ApJ...677L...5V,2011ApJ...739L..44D,2012MNRAS.422L..62C,2012ApJ...746..162N,2013MNRAS.428.1715H}) without significant changes in the slope (e.g.~\citealp{2012ApJ...746..162N}) or the scatter \citep{2014ApJ...788...28V}.

Properly interpreting these redshift-dependent evolutionary trends as a progenitor-descendant relation remains still elusive given the continuous mass build-up (e.g. \citealp{2013ApJ...777...18M,2013A&A...556A..55I}), the morphological transformations (e.g.~\citealp{2011arXiv1111.6993B}) and the evolution of the stellar populations (e.g~\citealp{2010ApJ...721..193P}) which make galaxies coming in and out of any given selection (\emph{progenitor bias} effect, e.g., \citealp{2012ApJ...746..162N,2013ApJ...773..112C,2015arXiv150102800S} and references therein).  As a matter of fact, a selection done at fixed stellar mass as usually done in several works, will necessarily be \emph{contaminated} by galaxies which grow in mass that will enter any given stellar mass bin. The level of contamination depends on the stellar mass selection. For massive galaxies ($\sim10^{11}$), the fraction of galaxies in the lowest redshift bin which are actually descendants of the galaxies at higher redshift ($z\sim2$) is less than 20\% \citep{2015MNRAS.450.3696M}. Therefore establishing evolutionary links is not  straightforward at all. The situation is even worse when passive galaxies are to be considered since quenching and mass growth both contribute to this progenitor bias effect. 

One popular solution has been to study the evolution of the number density of these compact objects (e.g. \citealp{2011MNRAS.415.3903T,2013ApJ...775..106C,2013ApJ...777..125P,2014ApJ...788...28V, 2015arXiv150104976D}), but the results are not always in agreement, specially at low redshifts where HST surveys probe a small area and also because of the multiple available definitions of \emph{compact galaxies}. As a matter of fact, some works do select only the most massive galaxies ($>10^{11}M_*/M_\odot$, e.g. \citealp{2011MNRAS.415.3903T}) while others select a wider stellar mass bin ($>10.5M_*/M_\odot$, e.g \citealp{2013ApJ...777..125P}).  On the other hand, there are papers using a fixed size threshold (a straight line in the mass-size plane, e.g. \citealp{2013ApJ...773..112C}) while others prefer a selection according to the slope of the mass-size relation (e.g \citealp{2014ApJ...788...28V,2013ApJ...775..106C,2013ApJ...765..104B}). On top of this, other parameters that could bias the results are environment and also the passive selection criteria (e.g \citealp{2013ApJ...777..125P,2010ApJ...721L..19V}). As a result, several authors do find a steep decrease of their abundances (e.g. \citealp{2013ApJ...775..106C}, \citealp{2014ApJ...788...28V}) while others tend to find a rather constant number (e.g. \citealp{2013ApJ...777..125P, 2013ApJ...773..112C, 2015arXiv150104976D}). 

Another option has been proposed based on selecting galaxies at fixed number density (e.g. \citealp{2010ApJ...709.1018V,2013ApJ...766...15P} and references therein), i.e. assuming the rank order is preserved at all epochs. This approach also implies some important assumptions such as neglecting the role of mergers or the scatter in the mass accretion histories and it is faced to the known uncertainties in the evolution of the massive end of the mass function (e.g.~\citealp{2014ApJ...797L..27S,2013MNRAS.436..697B}) and the continuous quenching happening at all cosmic epochs (e.g. \citealp{2010ApJ...721..193P}). Nevertheless, the latter approach can still provide some broad insights into 
the expected, average mass-growth of galaxies, thus allowing for a basic 
technique to observationally relate progenitors and descendants.
Moreover, the methodology has now been improved by including corrections to
the redshift-dependent number densities to account for mergers \citep{2014arXiv1412.3806P, 2014ApJ...794...65M, 2013ApJ...777L..10B} based on abundance matching.
It was also recently stressed that differences in the stellar mass function lead to consistent
results for the mass growth within $\sim0.25$ dex, at least for galaxies with $log(M_*/M_\odot) \leq11$   (e.g., \citealp{2014arXiv1412.3806P}). Globally, these empirical studies based on number conservation procedures tend to agree on a significant structural evolution, and confirm an important size growth of the average population. The growth seems to be driven by the addition of material in the outskirts of the galaxies (e.g. \citealp{2013ApJ...766...15P}) in what has been called an inside-out growth and interpreted as a minor merger driven growth through the tidal disruption of small companions falling into the central galaxy (e.g. \citealp{2009ApJ...699L.178N, 2010ApJ...725.2312O, 2012MNRAS.422.1714N, 2013MNRAS.428..109S} and references therein).

Despite the outstanding efforts made so far, it is still challenging to properly follow the evolutionary tracks
of especially the most massive galaxies. Number conservation-based approaches map today's red and dead early-type systems to \emph{progenitors} presenting a 
variety of morphologies and star-formation activities (e.g. \citealp{2008ApJ...688...48V,2011ApJ...743L..15V,2012MNRAS.427.1666B,2014arXiv1412.3806P, 2014arXiv1403.7524M}). It is thus very difficult to interpret the evolution of the overall population as a unique physical mechanism since multiple processes, such as morphological transformations or quenching can clearly contribute to move galaxies in the mass-size plane from one redshift bin to another.

By simultaneously following the evolution of the star formation activity
(quenching), morphological transformations, and mass build-up along the progenitor tracks identified through 
number-conserving techniques, one should be able to ideally separate the different contributors
to the average structural evolution.  


 All previous works lack however of a precise quantification of how the morphologies change and evolve among the progenitors of massive galaxies, mainly. The most significant effort has been probably done by~\cite{2012MNRAS.427.1666B} who made bulge-to-disk decompositions but just on one CANDELS field (UDS) and without quantification of irregularities. Another noticeable effort has been carried out by \cite{2013MNRAS.433.1185M},
who in the same UDS CANDELS field visually classified galaxies into discs, ellipticals and peculiar systems. 
They found significant evolution in the fractions of galaxies at a given visual classification as a function of redshift, though they did not
attempt to trace evolutionary tracks among galaxies at different epochs.

This is therefore the main new ingredient which motivates the present paper, in which we bring into the puzzle of massive galaxy formation detailed \emph{visual} like morphologies for a large sample of galaxies from all the five CANDELS fields. Combined with accurate stellar-masses and rest-frame colors as well as optical rest-frame structural parameters from the 3D-HST~\citep{2012ApJS..200...13B} and CANDELS~\citep{2011ApJS..197...36K,2011ApJS..197...35G} surveys, we revisit the evolutionary tracks of massive galaxies from $z\sim3$. 

The paper proceeds as follows. In section~\ref{sec:dataset}, we describe the dataset used as well as the main physical parameters derived (morphologies, structural parameters, SFRs etc..). In~\S~\ref{sec:nd} we describe the procedure to select the main progenitors and from~\S~\ref{sec:morpho_evol} to~\S~\ref{sec:struct} we describe the main results, namely the evolution of the morphologies, structures and star-formation properties. These results are discussed in~\S~\ref{sec:disc} and we provide a summary in~\S~\ref{sec:summ}.

Throughout the paper, we adopt a flat cosmology with $\Omega_M=0.3$, $\Omega_\Lambda=0.7$ and $H_0= 70$ $km.s^{-1}.Mpc^{-1}$ and we use magnitudes in the AB system.

\section{Dataset}
\label{sec:dataset}

\subsection{Parent sample}
 We select all galaxies in the F160W filter with F160W$<$24.5~mag (AB) in the 5 CANDELS fields (UDS, COSMOS, EGS, GOODS-S, GOODS-N). Our starting-point catalogs are the CANDELS public photometric catalogs for UDS \citep{2013ApJS..206...10G} and GOODS-S \citep{2013ApJS..207...24G} and preliminary CANDELS catalogs were used for COSMOS, EGS and GOODS-N (private communication). The magnitude cut is required to ensure reliable visual morphologies \citep{2014arXiv1401.2455K} and structural parameters \citep{2012ApJS..203...24V}  which are two key ingredients for the analysis presented in this work. As discussed in  \cite{2014ApJ...788...28V}, the magnitude cut results in a reasonable mass completeness of $log(M_*/M_\odot)\sim10$ at $z\sim3$ which is well beyond the mass limit required to follow the progenitors of massive galaxies as discussed in the following. Our results should thus not be significantly affected by incompleteness. The median redshift of the sample is $z\sim1.25$.

\subsection{Morphologies}
\label{sec:morphos}
Visual-like morphologies are taken form the deep-learning morphology catalog described in Huertas-Company et al. (2015a) on the 5 CANDELS fields UDS, COSMOS, GOODS-N, GOODS-S and EGS. The classification mimics the CANDELS visual classification scheme from \cite{2014arXiv1401.2455K} which is currently available in only one field. Morphologies are estimated using ConvNets, a specific artificial neural network topology that is inspired by the biological visual cortex (e.g.~\citealp{Fukushima80}) which is by far the most powerful image classifier up to date. When used for image recognition, convolutional neural networks consist of multiple layers of small neuron collections which look at small portions of the input image, also called receptive fields. The results of these collections are then tiled so that they overlap to obtain a better representation of the original image; this is repeated for every such layer. More details can be found in Huertas-Company et al. (2015a).

The algorithm is trained on GOODS-S for which visual classifications are publicly available and then applied to the other 4 fields. Following the CANDELS classification scheme, we associate to each galaxy 5 numbers - $f_{sph}$, $f_{disk}$, $f_{irr}$, $f_{PS}$, $f_{Unc}$ - measuring the frequency at which hypothetical classifiers would have flagged the galaxy as having a spheroid, having a disk, presenting an irregularity, being compact or point source and being unclassifiable/unclear.  As shown in Huertas-Company et al. (2015a), ConvNets are able to predict the fractions of votes given a galaxy image with a bias close to zero and $\sim10\%-15\%$ scatter. The fraction of miss-classifications is less than $1\%$.  We refer the reader to the aforementioned work for more details on how the morphologies are estimated. The important information to keep in mind for this work is that the classification is very close to a purely visual classification. We use only a classification in the H band (F160W) since the differences in the derived (broad) morphologies when using other filters are very small as shown in \cite{2014arXiv1401.2455K}.\\

We are interested in distinguishing bulge and disk growth so we use the 5 morphology estimators to define 5 main morphological classes as follows:

\begin{itemize}

\item \emph{pure bulges [SPH]:}~~$f_{sph}>2/3$ AND $f_{disk}<2/3$ AND $f_{irr}<1/10$
\item \emph{pure disks [DISK]:}~~$f_{sph}<2/3$ AND $f_{disk}>2/3$ AND $f_{irr}<1/10$
\item \emph{disk+sph [DISKSPH]:}~~$f_{sph}>2/3$ AND $f_{disk}>2/3$ AND $f_{irr}<1/10$
\item \emph{irregular disks [DISKIRR]:}~~$f_{disk}>2/3$ AND $f_{sph}<2/3$ AND $f_{irr}>1/10$
\item \emph{irregulars/mergers[IRR]:}~~$f_{disk}<2/3$ AND $f_{sph}<2/3$ AND $f_{irr}>1/10$

\end{itemize}

\begin{figure*}
\begin{center}
\includegraphics[width=0.99\textwidth]{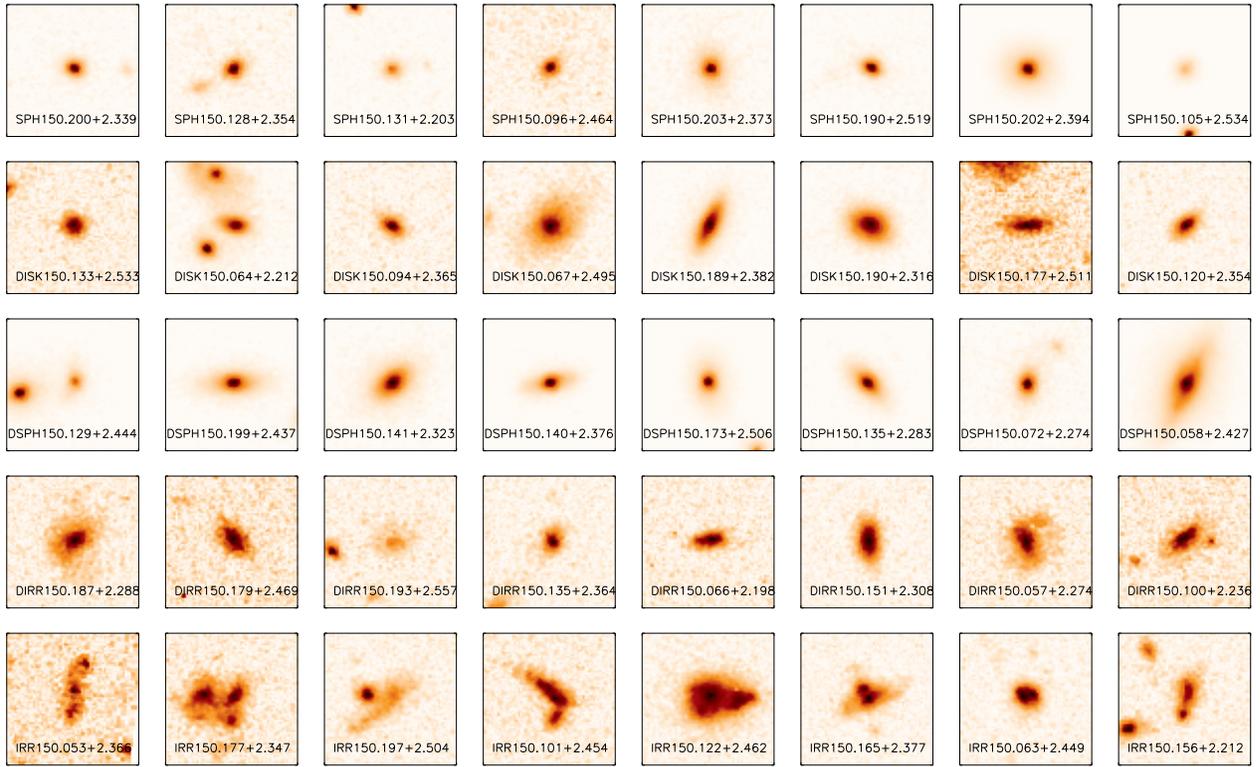} \\
\caption{Example stamps of the different morphological types defined in this work from the COSMOS field. From top to bottom, spheroids, disks, disk+spheroids, asymmetric disks and irregulars. Coordinates are indicated in each postage stamp.} 
\label{fig:stamps}
\end{center}
\end{figure*}

The classification accounts for the presence or not of a disk/bulge component as well as asymmetries in the light profile. The thresholds used are somehow arbitrary but have been calibrated through visual inspection to make sure that they result in different morphological classes (see also~\citealp{2014arXiv1401.2455K}). We emphasize that slight changes on the thresholds used to define the classes do not affect the main results of the paper. Figure~\ref{fig:stamps} shows some examples of the morphological classes defined that way. The SPH class contains galaxies fully dominated by the bulge component with little or no disk at all. The DISK class is made of galaxies in which the disk component dominates over the bulge. Between both classes, lies the DISKSPH class in which we put galaxies with no clear dominant component. Then we distinguish 2 types of irregulars: DISKIRR, i.e. disk dominated galaxies with some asymmetric features and IRR, which are irregular galaxies with no clear dominant disk component (including mergers).  These two last categories do contain all the variety of irregular systems usually observed in the high redshift universe (e.g. clumpy, chain, taphole etc..). The separation between the last 2 classes is however challenging (even for the human eye), since a diffuse light component can be easily interpreted as a disk. Therefore, even though we will consider the 2 classes separately in most of the plots, the reader should keep in mind that there can be significant overlap.  For the galaxies selected in this work (see section~\ref{sec:nd}), $>95\%$ of the population fits in one of the 5 defined classes. The remaining $\sim5\%$ contains either galaxies with rather high irregular, spheroid and disk morphologies simultaneously or unclassifiable objects.


\subsection{Stellar masses and star formation rates}

Photometric redshifts, stellar masses and star formation
rates (SFRs) are estimated from SED modeling as described in previous works by \cite{2011ApJ...742...96W,2012ApJ...753..114W} and \cite{2013ApJ...765..104B, 2014ApJ...791...52B}. We describe here the basic procedure and refer the reader to these works for more details. Photometric redshifts are estimated
from a variety of different codes available in the
literature which are then combined to improve the individual
performance. The technique is fully described in
\cite{2013ApJ...775...93D}. Based on the best available
redshifts (spectroscopic or photometric) we then estimate stellar masses and UV-based SFRs using FAST \citep{2009ApJ...705L..71K} assuming \cite{2003MNRAS.344.1000B} models, a \cite{2003PASP..115..763C} IMF, solar metallicity, exponentially declining star formation histories, and a
\cite{2000ApJ...533..682C} extinction law. Rest-frame magnitudes (U,V,J)
based on the best-fit redshifts and stellar templates
were computed using EAZY~\citep{2008ApJ...686.1503B}.

The final SFR used in this work combines IR-based and UV-based (from SED fitting) SFRs as described in \cite{2011ApJS..193...30B,2011ApJS..193...13B,2014ApJ...791...52B}. The method essentially relies on IR-based SFR
estimates for galaxies detected at mid- to far-IR wavelengths,
and SED-modeled SFRs for the rest. For IR-detected
galaxies the total SFRs, SFRIR+UV, are computed
from a combination of IR and rest-frame UV luminosities
(uncorrected for extinction) following \cite{1998ApJ...498..541K} and \cite{2005ApJ...625...23B}:

\begin{equation}
SFR_{UV+IR}=1.09\times10^{-10}(L_{IR}+3.3L_{2800}) [M_\odot.yr^{-1}]
\end{equation}

\subsection{Structural properties}

Structural properties (effective radii, Sersic indices and axis ratios) are taken from the public catalog released in~\cite{2012ApJS..203...24V}. Single Sersic 2D fits were performed to galaxies in CANDELS in the three infrared filters (f105,f125,f160) using galfit~\citep{2002AJ....124..266P}. The typical uncertainty on the parameters is less than $20\%$ for the magnitude cut applied in this work as clearly shown in~\cite{2012ApJS..203...24V}. \cite{2014MNRAS.443..874B} showed however that the total light profiles and sizes of massive galaxies at $z\sim0$ can significantly be affected by the background estimates. We do not expect a major impact of this effect in our sample at higher redshift though, where the contribution of the diffuse light around massive galaxies is less important. \cite{2014ApJ...788...28V} applied some corrections to the effective radii of passive and star-forming galaxies to measure them in a unique rest-frame band of $5000\AA$. Given that the corrections are very small and have little effect on the final measured structural evolution as discussed in the aforementioned work, we use here for simplicity the closest filter to the optical rest-frame band as done by \cite{2012ApJ...746..162N}. 


\section{Selecting the progenitors of massive galaxies}
\label{sec:nd}

\begin{figure*}
\begin{center}
\includegraphics[width=0.49\textwidth]{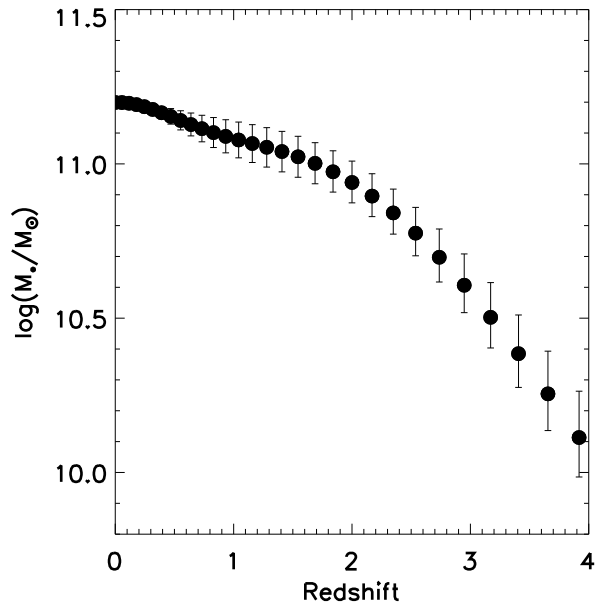} \\
\caption{Predicted mass growth of the progenitors of $log(M_*/M_\odot)=11.2$ galaxies from $z\sim4$ from the~\protect\cite{2013ApJ...777L..10B} model. Error bas show the errors on the median mass at a given redshift. } 
\label{fig:tracks}
\end{center}
\end{figure*}

One key issue when one tries to infer the evolution of individual galaxies is to actually link progenitors and descendants without being strongly affected by progenitor bias (e.g. \citealp{2013ApJ...773..112C,2014ApJ...786...89S,2015arXiv150102800S}). The stellar mass function (SMF) is known to significantly evolve from $z\sim3-4$ (e.g. \citealp{2008ApJ...675..234P, 2013ApJ...777...18M,2013A&A...556A..55I}) so a selection at fixed stellar mass will clearly be affected by new galaxies kicking in at lower redshifts as widely discussed in the recent literature. Also, a selection of only passive galaxies will be affected by the continuous quenching at all cosmic epochs. An alternative which is rapidly becoming very popular in the community is a selection at fixed number density (e.g. \citealp{2010ApJ...709.1018V, 2011ApJ...737L..31B,2013MNRAS.430.1051C,2013ApJ...766...15P}). At first level, this selection assumes that the ranking of galaxies is preserved at all redshifts and therefore deliberately ignores the impact of mergers and the scatter in the mass accretion histories~\citep{2013ApJ...777L..10B} which can lead to errors in the stellar mass growth of $d(logM_*)/dz\sim0.16$ dex (see also \citealp{2013ApJ...766...33L} for a comparison with SAM predictions leading to similar conclusions). To overcome this issue, \cite{2013ApJ...777L..10B} used abundance matching techniques to track the evolution of galaxies within their dark-matter haloes and apply a correction to this simple assumption. The model therefore accounts for number density evolution and is the one adopted in this work. Figure~\ref{fig:tracks} shows the stellar mass growth track for the progenitors of $\sim10^{11.2}$ galaxies from $z\sim4$ from the \cite{2013ApJ...777L..10B} model, assuming the stellar mass functions (SMFs) of \cite{2008MNRAS.388..945B, 2013ApJ...767...50M,2008ApJ...675..234P,2011MNRAS.413.2845M,2009ApJ...701.1765M,2010ApJ...725.1277M}. As recently shown by \cite{2014arXiv1412.3806P}, using different abundance matching assumptions (e.g. \citealp{2013MNRAS.428.3121M}) or different measured SMFs, leads to consistent results for the mass growth within $\sim0.25$ dex. The figure confirms that massive galaxies grow by a factor of 2 in stellar mass from $z\sim2$ and a factor of $\sim5$ from $z\sim3$, so that the typical stellar mass of the progenitors of  $\sim10^{11.2}M_*/M_\odot$ galaxies is $10^{10.5}$ at $z\sim3$ and $10^{10.0}$ at $z\sim4$. This mass growth track includes mergers, which occur at a rate of $\sim1.2$ major (1:4) mergers/galaxy between $z\sim3$ and $z\sim0.5$ (see section~\ref{sec:mergers} for a detailed discussion on the effect of mergers). Since our sample is mass complete down to $10.0$ from $z\sim3$, a selection along the progenitors should not be affected by incompleteness.  As described in \cite{2013ApJ...766...15P}, we select galaxies along the growth track by picking galaxies in a given redshift bin within a narrow stellar mass bin of $0.3$ dex around the corresponding mass for that redshift. As also discussed in \cite{2014arXiv1412.3806P}, this stellar mass bin is a reasonable trade off to account for the different predictions of different methods/SMFs and the scatter in the mass accretion histories. Table~\ref{tbl:tbl_summ} summarizes the main properties of the selected sample at different redshifts. The redshift bins are selected to keep a comparable number of objects in each bin ($\sim400$, except for the first and last bins) and as a tradeoff between comoving volume ($\sim 3.10^{5}$ $Mpc^{3}$) and lookback time ($0.5-1$ Gyr).

\begin{table*}
\begin{center}
\begin{tabular}{c c  c c c c c c c c c}
\hline
        \hline
        &  &  &  & N & \% & \% & \% & \% & \% & \%  \\
        Redshift   & c. vol. & time & $log(M_*/M_\odot)$  & TOT & SPH & DISK& DISKSPH& DISKIRR& IRR & OTHER   \\
        & ($Mpc^{3}$)& ($Gyr$) & & & & & & & & \\
         (1)& (2)& (3) & (4)&(5) &(6) &(7) &(8) & (9)& (10)&(11) \\
        \hline
        \hline

0.10$<z<$0.60&   $5.91\times10^{4}$& 4.27&11.17$\pm0.3$& 76& 34& 18& 34&  1&  7&  1\\
0.60$<z<$1.10&   $1.86\times10^{5}$& 2.31&11.10$\pm0.3$&455& 28& 23& 35&  5&  4&  2\\
1.10$<z<$1.60&   $2.74\times10^{5}$& 1.36&11.05$\pm0.3$&416& 29& 18& 24& 14&  7&  4\\
1.60$<z<$2.10&  $3.15\times10^{5}$ & 0.87&10.97$\pm0.3$&482& 30& 17&  9& 25& 11&  4\\
2.10$<z<$2.60&   $3.27\times10^{5}$& 0.59&10.84$\pm0.3$&319& 26&  6&  2& 31& 24&  8\\
2.60$<z<$3.00&   $2.60\times10^{5}$& 0.35&10.67$\pm0.3$&157& 14&  6&  1& 41& 28&  7\\
  \hline
\end{tabular}
\caption{Summary of selected objects. (1) redshift range, (2) comoving volume probed in the corresponding redshift range considering the CANDELS area, (3) lookback time interval, (4) stellar mass range, (5) total number of objects, (6) number of spheroids, (7) number of disks, (8) number of disk+spheroids, (9) number of irregular disks, (10) number of irregulars, (11) remaining galaxies which include unclassified, disk+irr+spheroids and sph+irr }
\label{tbl:tbl_summ}
\end{center}
\end{table*}

\section{Morphological evolution}
\label{sec:morpho_evol}
\begin{figure*}
\begin{center}
$\begin{array}{c}
\includegraphics[width=0.49\textwidth]{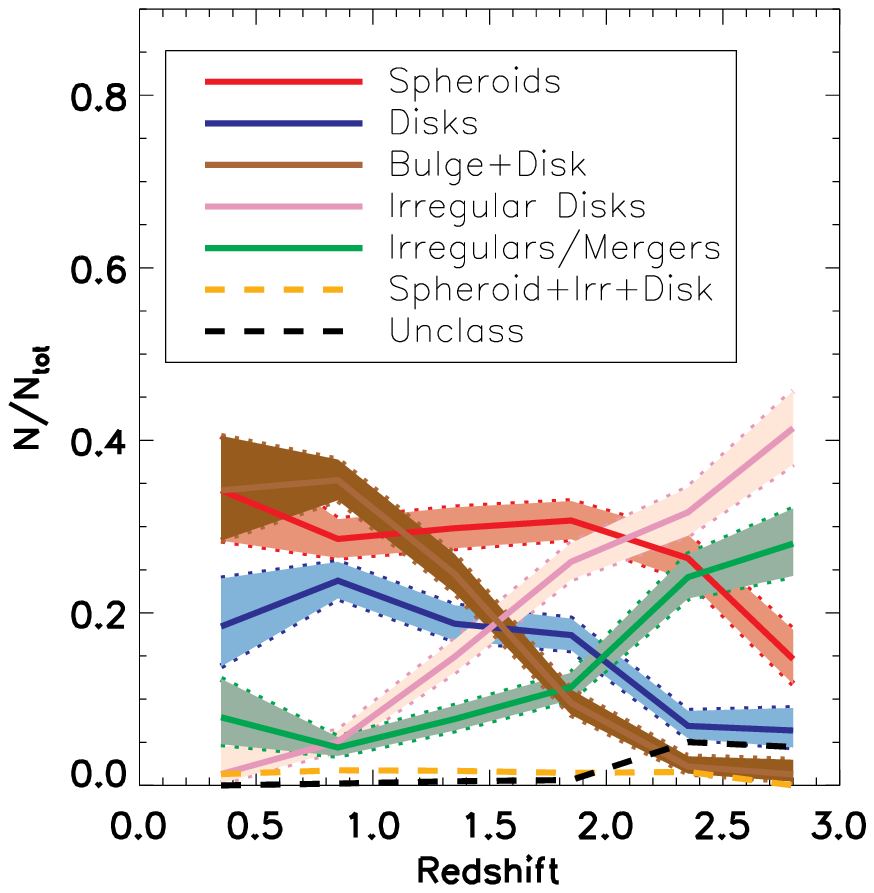} \\
\includegraphics[width=0.49\textwidth]{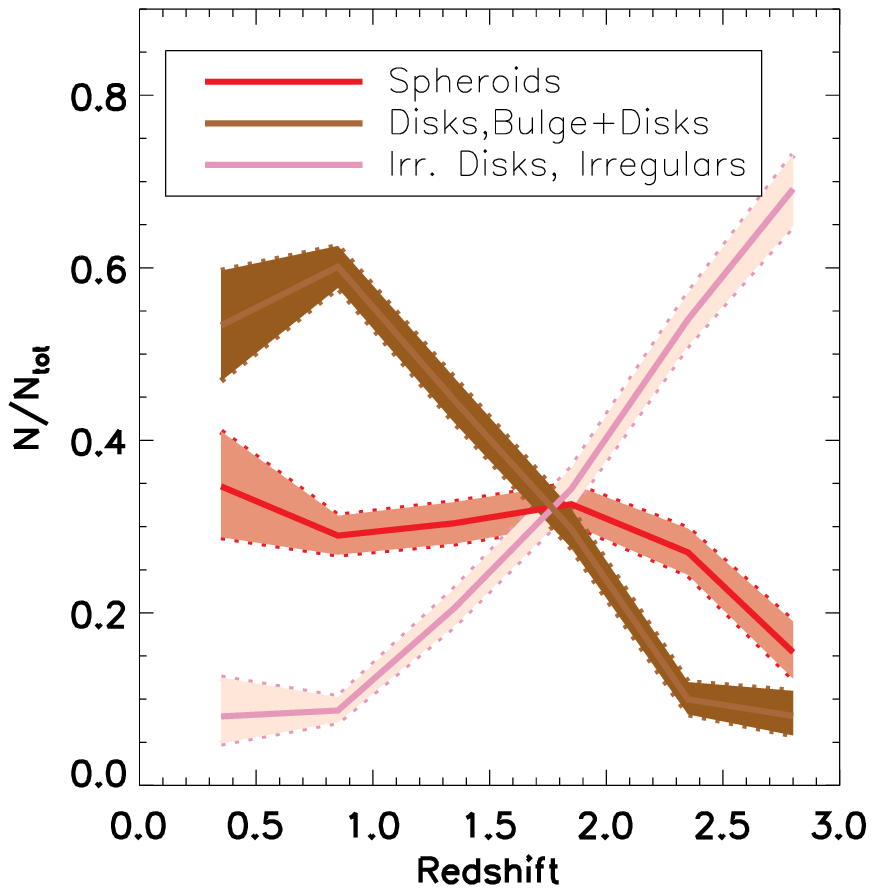} \\
\end{array}$
\caption{Evolution of the relative abundance of different morphological types as labeled between $z\sim0$ and $z\sim4$ along the mass growth tracks from~\protect\cite{2013ApJ...777L..10B} (see text for details). The shaded regions indicate the $1\sigma$ error on the fractions computed following~\protect\cite{1986ApJ...303..336G} (see Section 3 for binomial statistics; see also \protect\citealp{2009ApJ...690...42M}). The top panel shows all the morphological types defined in~\ref{sec:morphos}. In the bottom panel, all the irregulars are combined in one class and all the \emph{disky} galaxies in another one. } 

\label{fig:morpho_fracs}
\end{center}
\end{figure*}

\begin{figure*}
\begin{center}
\includegraphics[width=0.94\textwidth]{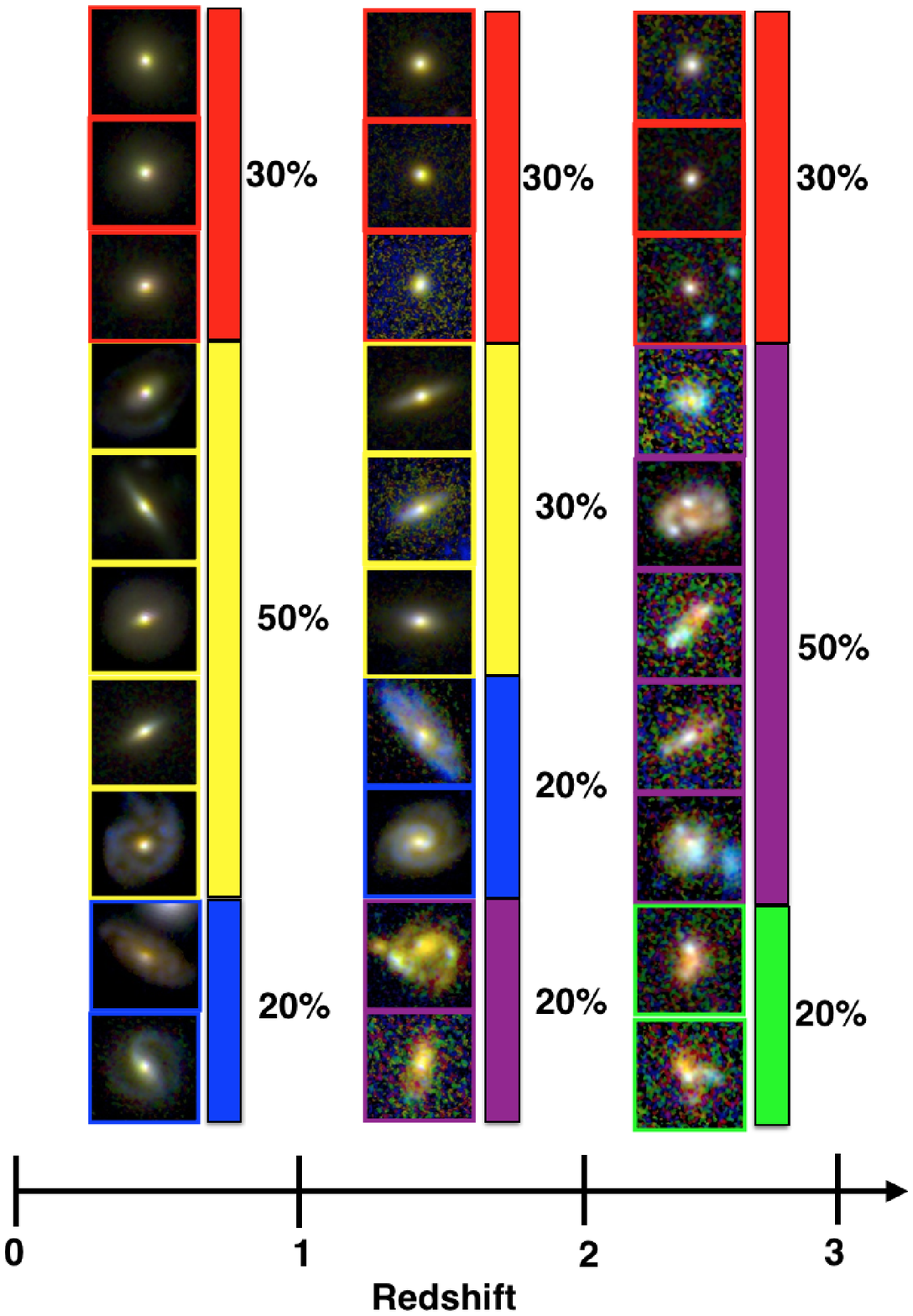} \\
\caption{Color stamps illustrating the evolution of the relative abundance of each morphological type along the main progenitors. Each column shows a different redshift bin, $0<z<1$, $1<z<2$ and $2<z<3$ from lest to right. The stamps are roughly built in the same rest-frame color using $f814$, $f105$, $f125$ or $f160$ depending on the considered redshift. All stamps are normalized to the maximum pixel value. } 
\label{fig:hubble_seq}
\end{center}
\end{figure*}

Figure~\ref{fig:morpho_fracs} shows the evolution of the relative abundance of the different morphological types defined in~\S~\ref{sec:morphos} selected along the mass growth track from~figure~\ref{fig:tracks} in 0.3 dex bins. The  plot confirms the strong morphological evolution experienced by the population of massive galaxies between $z\sim3$ and $z\sim1$ essentially (see also e.g. \citealp{2011ApJ...743L..15V,2012MNRAS.427.1666B,2013MNRAS.433.1185M}). About $60-80\%$ of the progenitors of massive galaxies at $z\sim3$ were irregular disks ($\sim40-50\%$) and mergers/irregulars ($\sim20-30\%$), while the population at $z<1.0$ is made at $80\%-90\%$ by pure spheroids and galaxies with a classic bulge+disk structure. Below $z\sim1$, the well-known massive end of the Hubble sequence seems to be in place in terms of morphological mixing. Figure~\ref{fig:hubble_seq} illustrates this morphological transformation with some example color stamps. Hence, considering all the progenitors of massive galaxies as an homogeneous family of objects when trying to infer their structural evolution, necessarily ignores the striking diversity of morphologies and the effect of morphological transformations.

The inspection of the evolution for each morphology individually reveals some interesting trends. The fraction of \emph{pure} spheroids is in fact roughly constant with redshift and represents about $\sim30\%$ of the population of massive galaxies at all epochs (only a slight decreasing trend is observed at $z>2.5$). Most of the evolution is observed in the bulge+disk and the irregular disks populations which present more or less symmetric trends as clearly shown in the bottom panel of figure~\ref{fig:morpho_fracs}. The latter goes from $\sim60\%$ of the population of massive galaxies at $z\sim3$ to roughly $\sim5\%$ at $z\sim0.2$. This decrease is mirrored by the increase of the disk and bulge+disk populations which are almost inexistent at $z=3$ and represent $50\%$ of the galaxy population at low redshift. These trends suggest that most of the morphological transformations going on in the progenitors of massive galaxies go into one single direction, i.e from irregular/clumpy disks to more regular bulge+disk galaxies while the population of pure spheroids remains unaltered from $z\sim2.5$ and might follow an independent evolutionary track. 

The result might be an indication of two independent channels for bulge growth in massive galaxies acting at very different timescales. Around $\sim30\%$ of the population of massive galaxies at $z\sim0$ was already made of bulges at $z\sim2.5$ with probably an early (\emph{monolithic}) fast collapse. The other half, however, have clearly a disk component and seem to appear gradually from $z\sim3$ and $z\sim1$ ($\sim 3Gyr$) through the morphological transformation of clumpy-irregular disks possibly through the migration of clumps and stabilization of the disks (e.g. \citealp{2014ApJ...780...57B}).

\section{Star formation}
\label{sec:SFR}
\begin{figure*}
\begin{center}
$\begin{array}{c }
\includegraphics[width=0.94\textwidth]{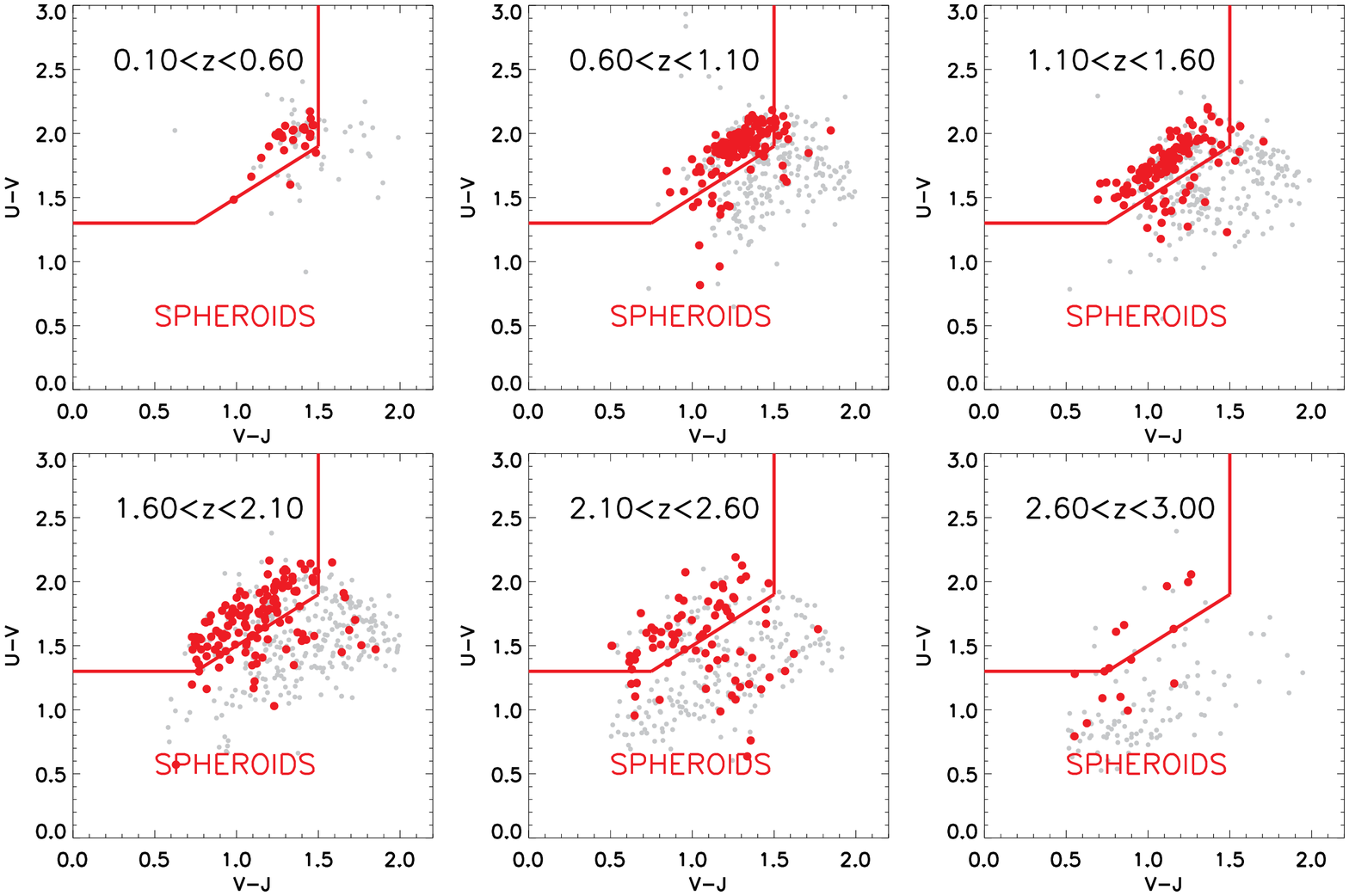} \\
\includegraphics[width=0.94\textwidth]{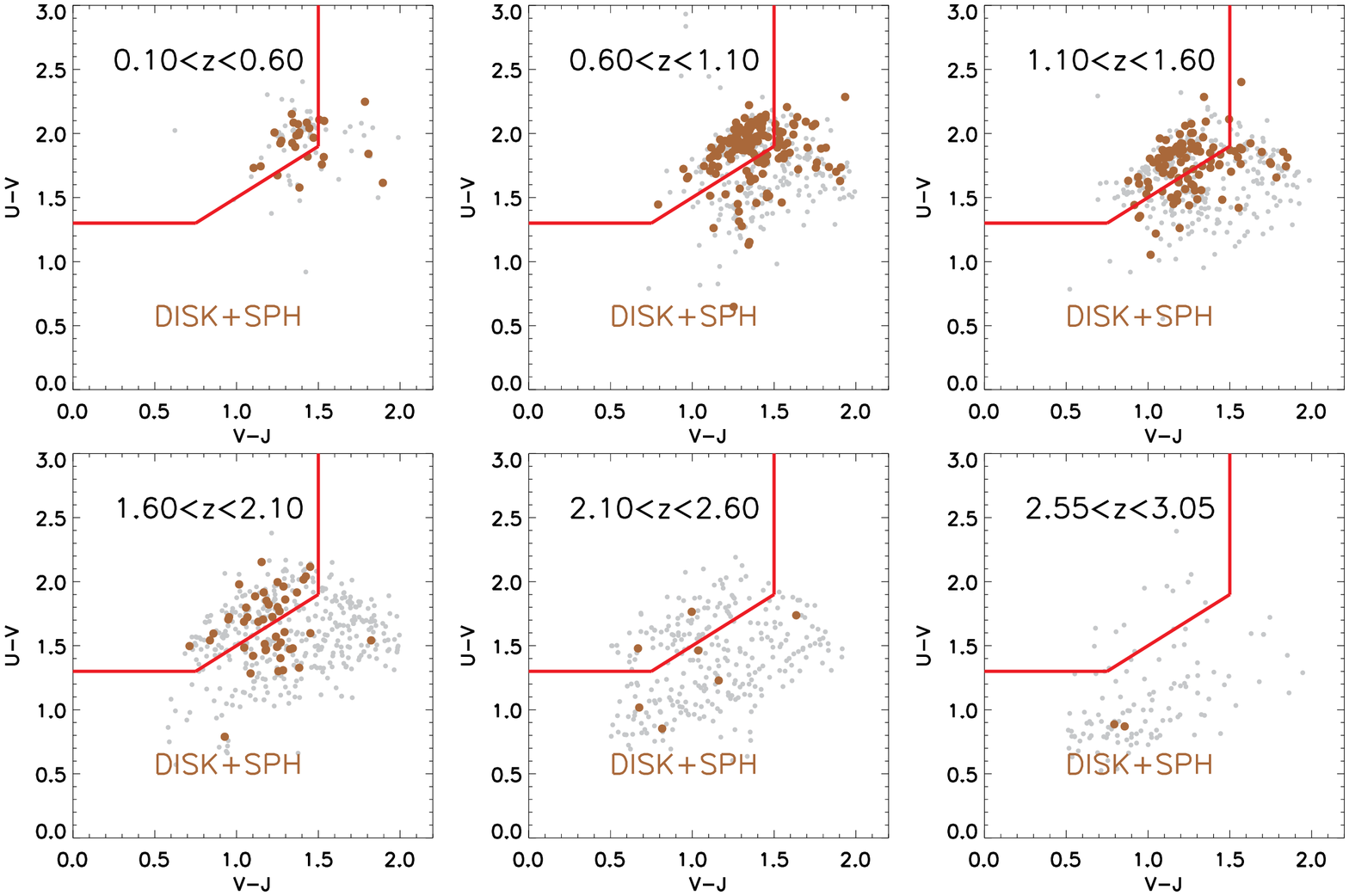} 
\end{array}$
\caption{Rest-frame UVJ plane for spheroids (top panels) and disk+sph (bottom panels) at different redshifts as labelled. The red lines indicate the quiescent region as defined by~\protect\cite{2012ApJ...754L..29W} and  gray points are all galaxies in the corresponding redshift/mass bin.} 
\label{fig:color_color_1}
\end{center}
\end{figure*}

\begin{figure*}
\begin{center}
$\begin{array}{c }
\includegraphics[width=0.94\textwidth]{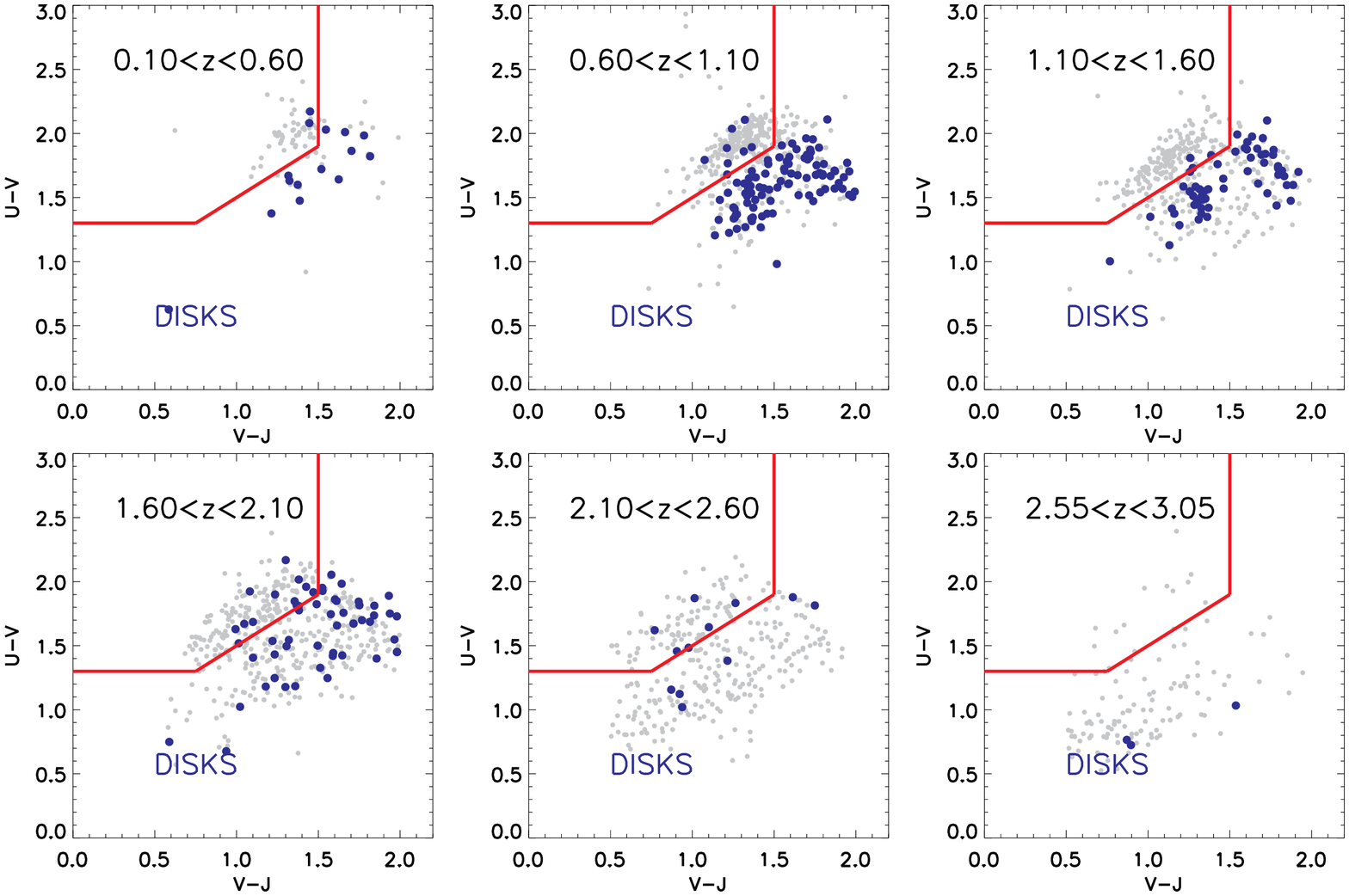} \\
\includegraphics[width=0.94\textwidth]{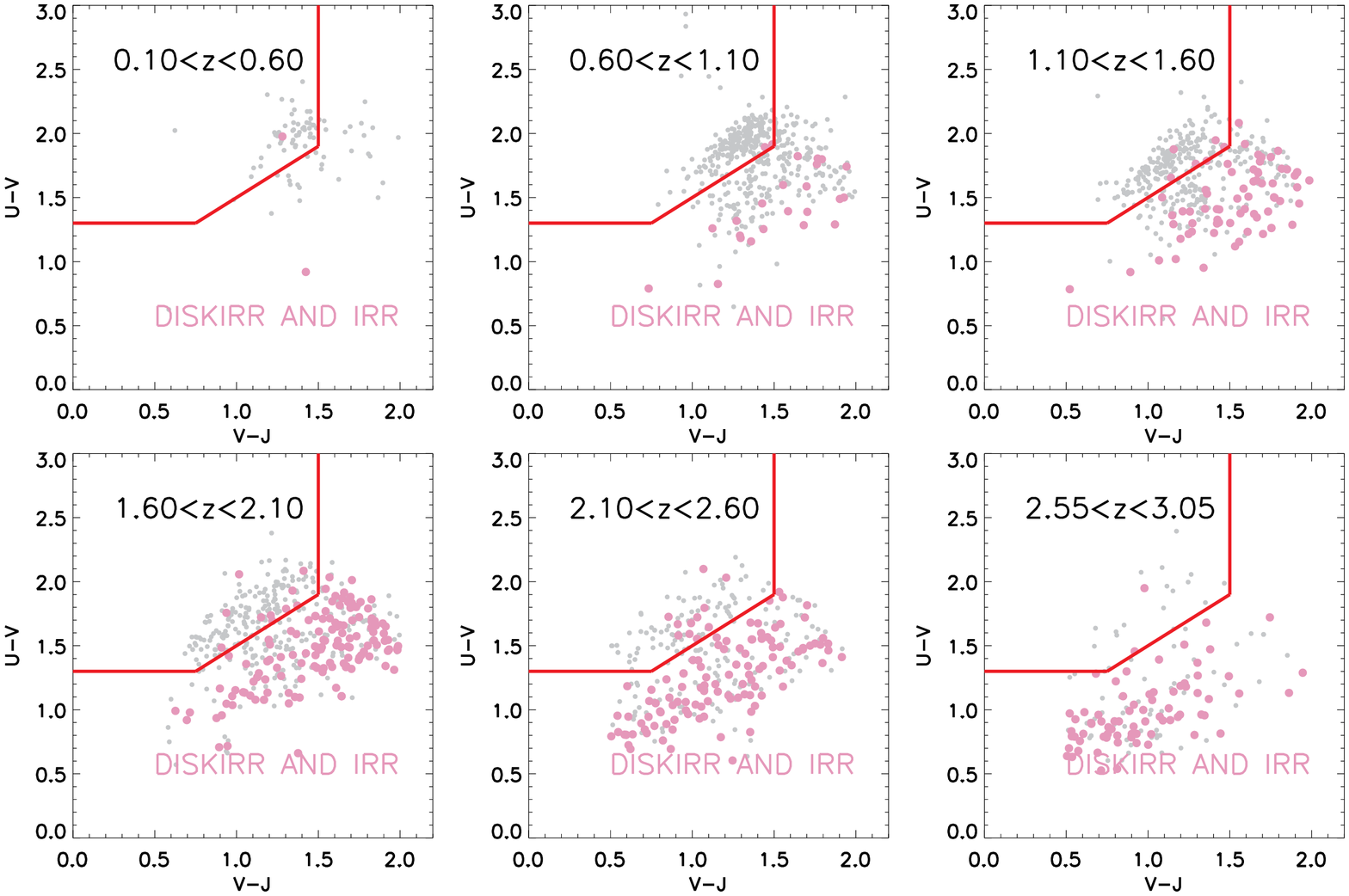} 
\end{array}$
\caption{Rest-frame UVJ plane for disks (top panels) and disk irregulars (bottom panels) at different redshifts as labelled. The red lines indicate the quiescent region as defined by~\protect\cite{2012ApJ...754L..29W} and  gray points are all galaxies in the corresponding redshift/mass bin. } 
\label{fig:color_color_2}
\end{center}
\end{figure*}

\begin{figure*}
\begin{center}
$\begin{array}{c c}
\includegraphics[width=0.50\textwidth]{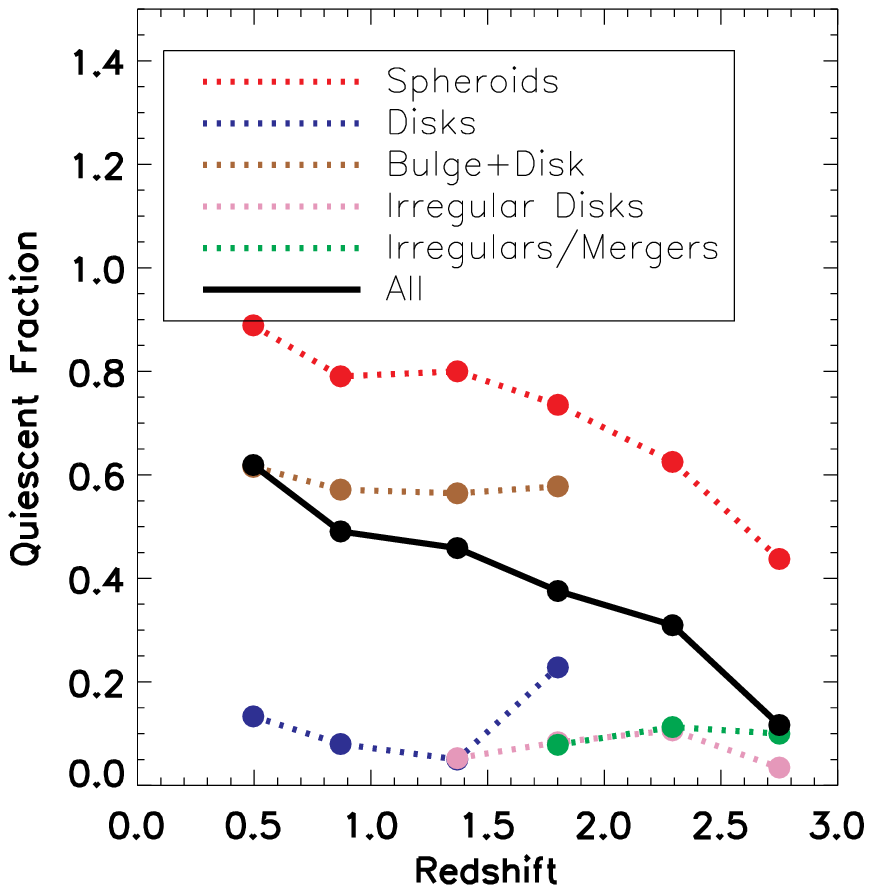} & \includegraphics[width=0.50\textwidth]{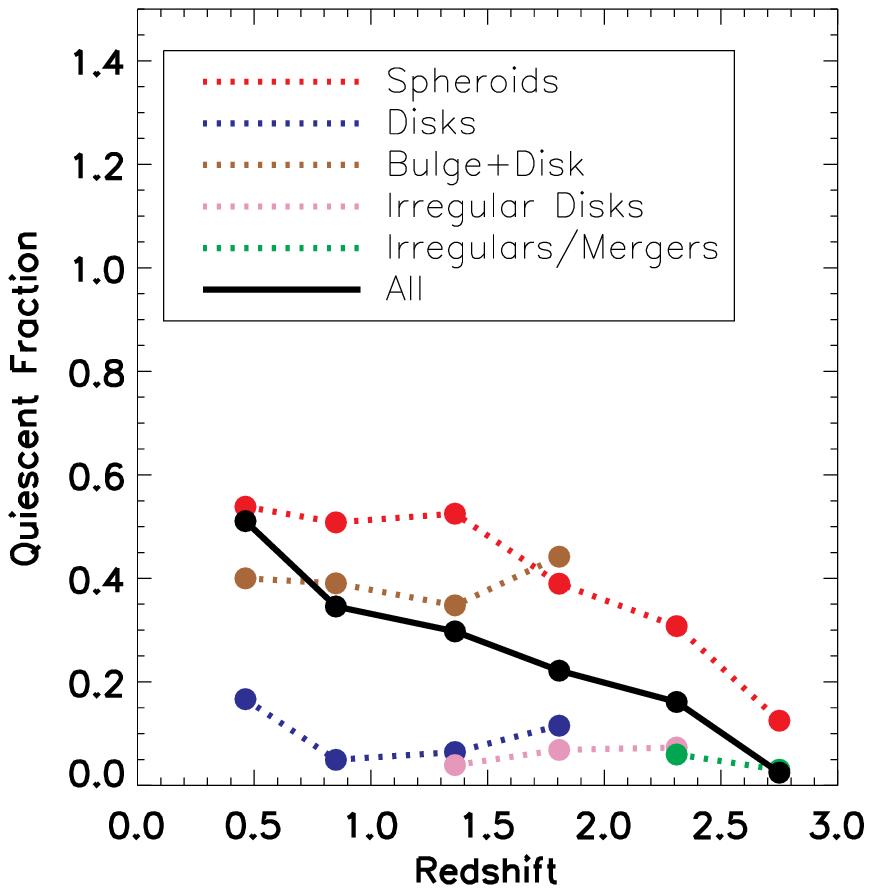}\\
\end{array}$
\caption{Quiescent fraction as a function of redshfit for different morphological types as labelled. In the left panel, the quiescent fraction is computed using the UVJ plane while in the right panel a threshold in sSFR ($log(sSFR)<-1.5$ $[Gyr^{-1}]$) is used.  The trends are the same, but the absolute number of objects considered as passive changes depending on the definition used.} 
\label{fig:Q_frac}
\end{center}
\end{figure*}

We now explore how the stellar populations, star formation rates and gas fractions evolve for each morphological type. Figures~\ref{fig:color_color_1} to~\ref{fig:color_color_2} show the evolution of the UVJ planes for different morphological types. Objects with different morphologies clearly populate different regions of the color-color plane as expected. Disk dominated galaxies (disks and irregular disks) tend to populate the star-forming region at all redshifts while pure spheroids are more concentrated towards the quiescent zone. Disk+spheroid galaxies lie between both regions. This confirms, that while a separation between passive and star-forming galaxies, as for example done by \cite{2014ApJ...788...28V}, is clearly correlated with the morphology, it will not result in a clean separation of the morphological types and will mix bulges and disks. This is better seen in the left panel of figure~\ref{fig:Q_frac} which shows the quiescent fraction for different morphologies, where quiescent galaxies are selected using the UVJ plane (red box in figures~~\ref{fig:color_color_1} to~\ref{fig:color_color_2}). The average population is clearly quenched between $z\sim3$ and $z\sim0.5$ with the quiescent fraction rising from $\sim20\%$ at $z\sim3$ to $\sim80\%$ at $z\sim0$, in agreement with the findings of \cite{2013ApJ...766...15P} and \cite{2014arXiv1412.3806P}. However, $90\%$ of the disks and irregular disks are star-forming at all redshifts not showing any significant increase in the number of passive galaxies. On a  similar vein, bulge+disk galaxies have a rather constant quiescent fraction at all redshifts, close to $60\%$. The spheroid population however shows a clear increase going from a passive fraction of $60\%$ at $z\sim3$ to almost $90\%$ at $z\sim0$. Given that the number density of spheroids remains roughly constant in the redshift range probed, this trend can be easily  interpreted as the same galaxies being quenched (within the limits of the abundance matching based selection). The increasing quiescent fraction observed for the overall population could then be explained as a combination of morphological transformations of disk-irregular galaxies becoming disk+spheroids (as suggested by figure~\ref{fig:morpho_fracs}) and spheroids being individually quenched. 

\begin{figure*}
\begin{center}
$\begin{array}{c }
\includegraphics[width=0.40\textwidth]{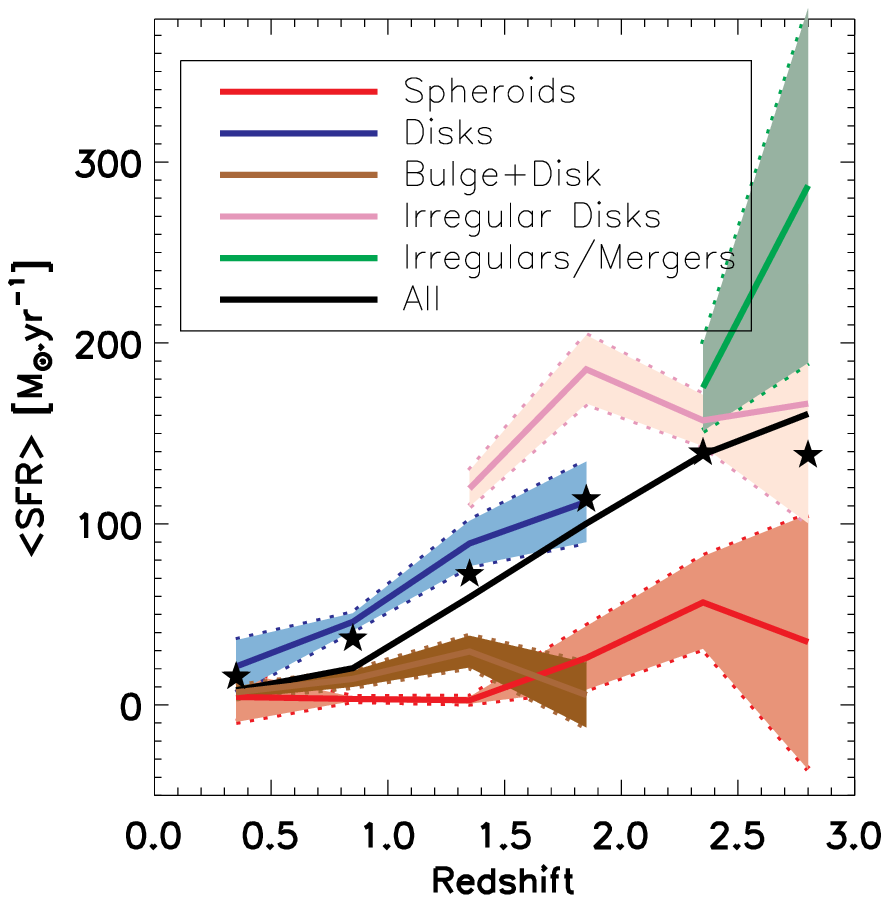} \\
\includegraphics[width=0.40\textwidth]{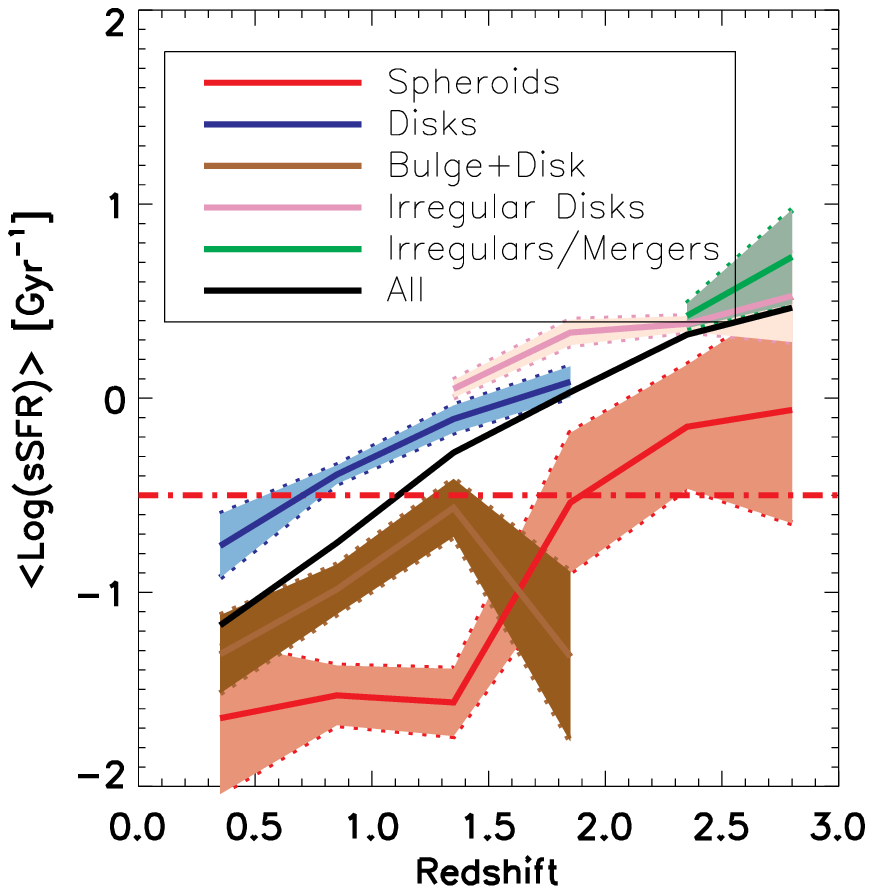} 
\end{array}$
\caption{Median SFR (top) and sSFR (bottom) for different morphological types. Red: spheroids. yellow: disk+sph, blue:disks, violet:disk+irr, green:irr. The shaded regions show the $1\sigma$ uncertainty estimated through bootstrapping.  Black stars in the top panel show the position of main sequence galaxies at a given redshift according to the measurement of~\protect\cite{2012ApJ...754L..29W}. The red dashed-dotted line in the bottom panel shows the limit used by~\protect\cite{2013ApJ...765..104B} to define star-forming and quiescent galaxies. We only plot morphologies which represent at least $10\%$ of the total population at a given epoch.} 
\label{fig:sfr}
\end{center}
\end{figure*}

A similar conclusion arises from figure~\ref{fig:sfr} in which we plot the median star-formation rate (SFR) and specific star-formation rate (sSFR) for all morphologies. Different morphologies form stars at very different rates at all epochs, ranging from several hundreds of solar masses per year for the irregular and irregular disks to a few tens for spheroids. Generally speaking, objects with a significant bulge component tend to lie below the star formation main sequence at all redshifts (\citealp{2012ApJ...754L..29W} shown with stars in fig.~\ref{fig:sfr}). The overall trend (black line in figure~\ref{fig:sfr}) is however a clear decrease of the SFR irrespective of the morphological type as predicted by several models (e.g. \citealp{2008MNRAS.383..119D,2010ApJ...721..193P}) and in agreement with the evolution of the star formation main sequence \citep{2012ApJ...754L..29W}. \\

Spheroids had a modest (compared to the average main sequence at that epoch, i.e. \citealp{2012ApJ...754L..29W}) star formation activity already at $z\sim2-3$ ($SFR\sim50M_\odot .yr^{-1}$) suggesting again that the formation of their stellar content occurred at earlier epochs and that they are in the process of quenching, i.e their star formation rate at $z\sim0.5$ is almost 0.  We do observe however a significant increase of the average sSFR and above $z\sim2$ it becomes larger than the threshold used by \cite{2013ApJ...765..104B} to define quiescent galaxies ($log(sSFR[Gyr^{-1}])=-0.5$).  This increase is also accompanied by an increase of the scatter as also shown in figure~\ref{fig:sfr}. At $z>2$, a significant fraction of spheroids are therefore actively forming stars, at similar rates than $z\sim1$ main sequence disks (see also \citealp{2013ApJ...765..104B}).

Clumpy disks have rather high SFRs ($>100M_\odot.yr^{-1}$) at all epochs in which they are still abundant ($z>1-1.5$) as well as disk dominated galaxies, roughly lying in the main sequence. Bulge+disk galaxies from roughly $50M_\odot.yr^{-1}$ departing from the star-formation main sequence. This suggests that, while the transition from irregular disks to disk dominated systems appears to be smooth without a big impact on the star formation activity, the morphological transformation between irregular and bulge+disk (i.e. the emergence of the bulge component) has to be accompanied by a decrease of their star formation activity and a departure from the main sequence. In other words, the emergence of the bulge and the stabilization of the disk in these objects tends to decrease significantly their SFR. This behavior is in line with the predictions of several numerical simulations (e.g. \citealp{2009ApJ...707..250M}) which predict that the growth of a bulge in a turbulent disk can be sufficient to stabilize the gas disk and quench star formation (\emph{morphological quenching}). Another possibility is that the quenching which seems to follow the growth of the bulge is driven by the effects of a super massive black hole in the growing bulge (e.g.~\citealp{1998A&A...331L...1S,2004ApJ...600..580G}). 

To follow up on this idea, we look at the gas fractions through the existing correlation between the surface density of the star formation rate and the cold gas through the Kennicut-Schmidt law \citep{1959ApJ...129..243S,1998ApJ...498..541K}. As done in \cite{2013MNRAS.430.1051C} and~\cite{2014arXiv1412.3806P} we use the following relation to infer gas masses:
\begin{equation}
\frac{M_{gas}}{6.8\times10^8M_\odot}=\left (\frac{SFR}{1M_\odot.yr^{-1}}\right )^{5/7}\left (\frac{R_e}{1kpc}\right )^{4/7}
\end{equation}

which is then used to estimate the gas fraction as:

\begin{equation}
f_{gas}=\frac{M_{gas}}{M_{gas}+M_*}
\end{equation}

\begin{figure*}
\begin{center}
\includegraphics[width=0.44\textwidth]{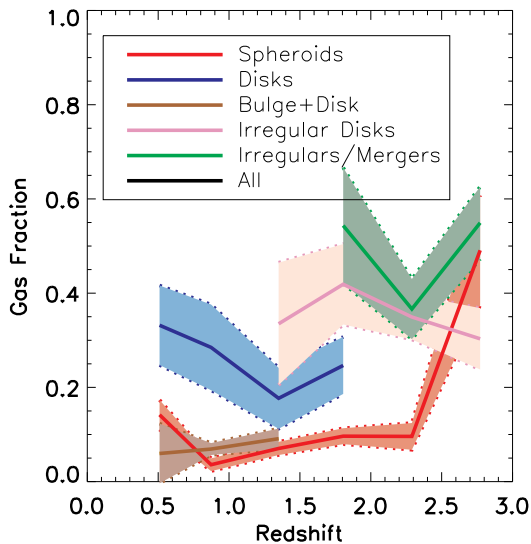} \\
\caption{Median gas fractions or different morphologies. The black solid line show the average population while the different colors show different morphologies: red:spheroids, yellow, disk+sph, blue:disks. violet: disk irregulars and green irregulars. The shaded regions show the $1-\sigma$ errors on the mean estimated through bootstrapping. We only plot morphologies which represent at least $10\%$ of the total population at a given epoch.} 
\label{fig:gas_frac}
\end{center}
\end{figure*}

Figure~\ref{fig:gas_frac} shows the evolution of the inferred gas fractions for different morphologies. The average gas fraction decreases monotonically with redshift from a value of $\sim60\%$ to $\sim20\%$ at $z\sim0.5$ as already reported in \cite{2014arXiv1412.3806P} for a slightly less massive sample. The trends differ significantly for different morphologies though. Spheroids, tend to have low gas fraction ($\sim10\%$) at least from $z\sim2$ while, irregulars and disk-irregulars keep high gas mass fractions ($50-60\%$). 

The increase of the gas fraction of spheroids at $z>2.5$, even with the large uncertainties, is line with the idea of these objects rapidly assembling at these epochs and consuming their gas reservoirs. On the other hand, the decrease of the gas content in disk galaxies, is again tightly linked with the emergence of the bulge component. While the decrease is rather smooth when no significant bulge is built, it becomes more dramatic for galaxies with a more predominant bulge (decreasing from $\sim40\%$ to $\sim20\%$). 

\section{Structure}
\label{sec:struct}

\begin{figure*}
\begin{center}
$\begin{array}{c}
\includegraphics[width=0.40\textwidth]{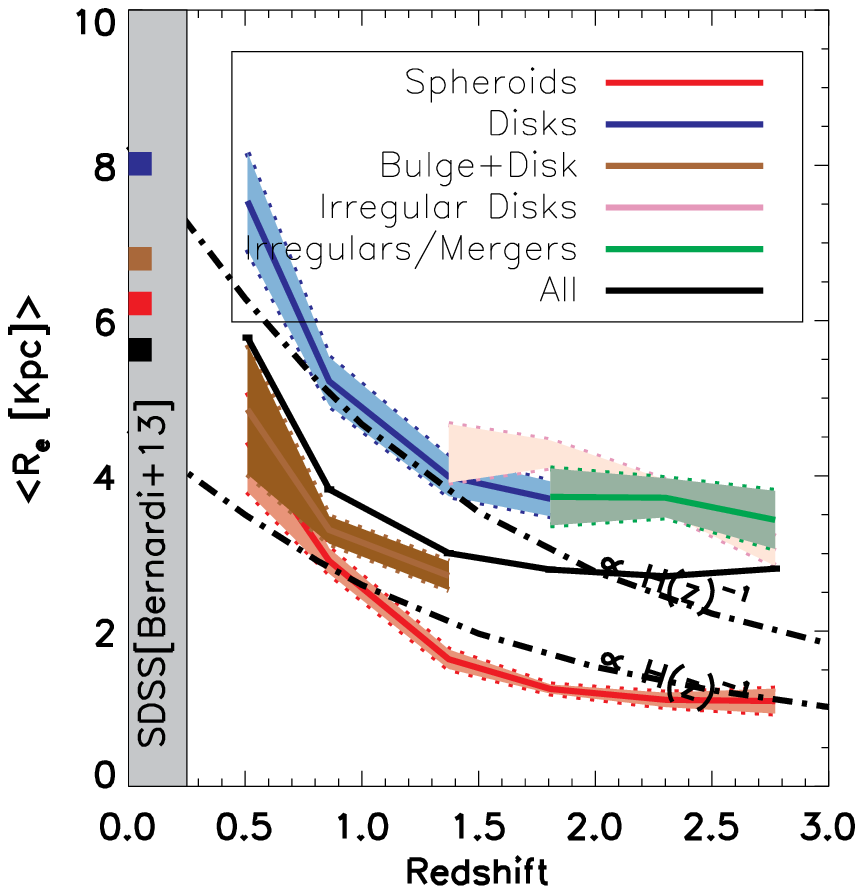} \\ 
\includegraphics[width=0.40\textwidth]{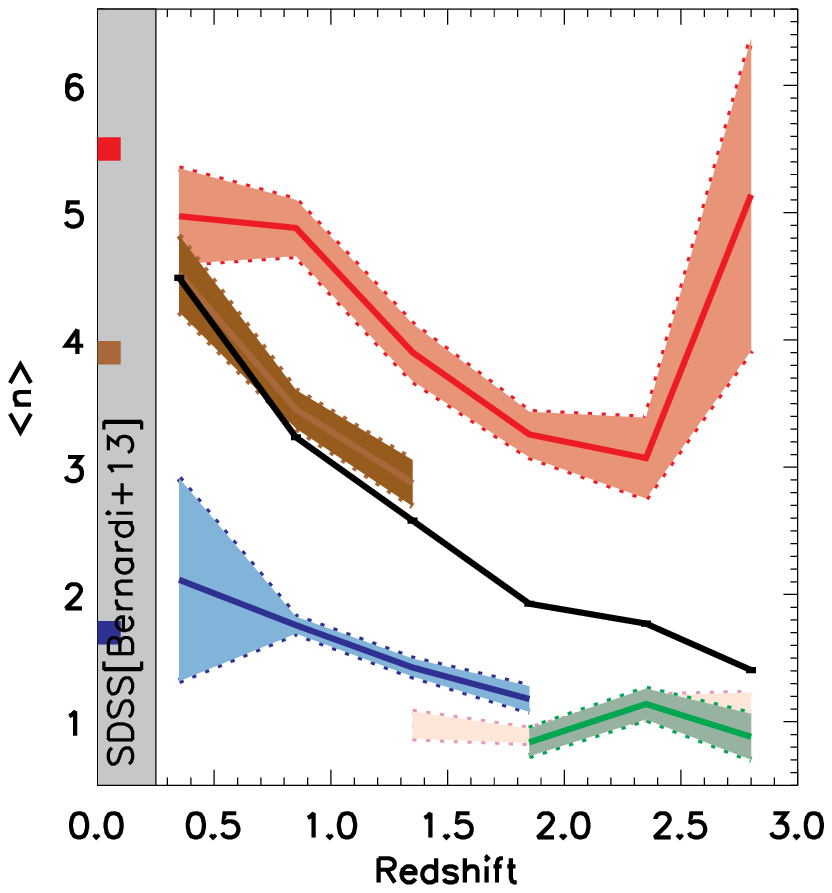} \\
 \includegraphics[width=0.40\textwidth]{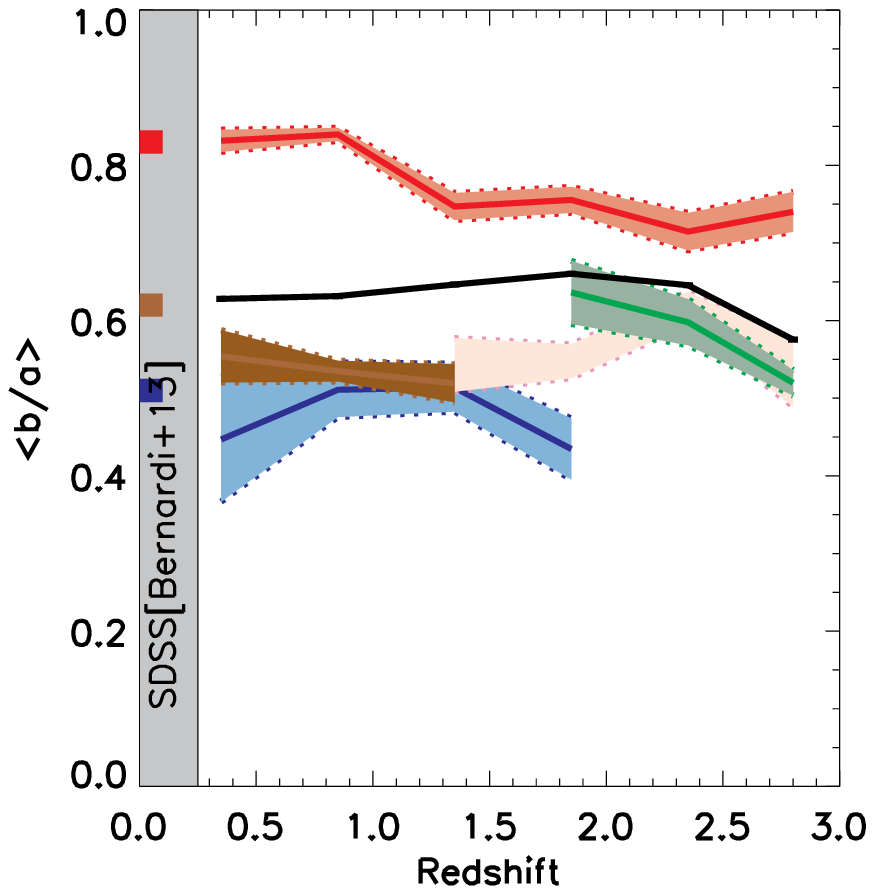}\\

\end{array}$
\caption{Evolution of the median effective radius (left panel), Sersic index (middle panel) and axis ratio (right panel) for different morphologies. The red, yellow, blue, magenta and green lines show spheroids, bulge+disks, disks, irregular disks and irregulars respectively. The black line shows all galaxies irrespectively of their morphology. Error bars are estimated through bootstrapping. The squares show the values at $z\sim0$ from~\protect\cite{2014MNRAS.443..874B} and the dashed-dotted black lines in the left panel show the relation $H(z)^{-1}$ normalized at the spheroids and disks $z\sim0.8$ value.} 
\label{fig:med_props}
\end{center}
\end{figure*}

We now move on to the study of the evolution of the structural properties of the different morphological types. Figure~\ref{fig:med_props} shows the evolution of the effective radii, Sersic indices and axis-ratios. There is an average size increase of a factor of $\sim2$ from $z\sim3$, as already pointed out by many works (e.g~\citealp{2005ApJ...626..680D,2006ApJ...650...18T,2008ApJ...687L..61B,2008ApJ...677L...5V,2011ApJ...739L..44D,2012MNRAS.422L..62C,2012ApJ...746..162N,2013MNRAS.428.1715H}). We do clearly find two regimes in the size growth as also discussed by \cite{2013ApJ...766...15P} for a similar selection. From $z\sim3$ to $z\sim1.5$, the average size of the whole population remains roughly constant and starts a sharp increase from $z\sim1.5$ to $z\sim0$. Recall that this differers from other works selected at fixed stellar mass (e.g.~\citealp{2012ApJ...746..162N}) because the selection is different. Adding the information of the morphological evolution discussed in~\S~\ref{sec:morphos}, these two phases in the structural evolution are better explained.  
From $z\sim3$ to $z\sim1.5$ there is a rapid morphological transformation of irregular disks into bulge+disk systems. Even though irregular disks are rapidly increasing their effective radii, their number density is also decreasing fast to be transformed into bulge+disk galaxies, which results in a decrease of the effective-radius because of the mass going into the central parts of the galaxy to build the bulge. As a result, both effects compensate to produce a flat size evolution. From $z\sim1.5$, the morphological mixing remains roughly constant and the average growth reflects simply the growth of the different morphological types. Interestingly, all dominant morphologies (spheroids, disks and disk+spheroids) at these redshifts do grow but the growth rate is different. While spheroids increase their effective radii by a factor of $\sim3$ ($\sim 5$ from $z\sim3$), disks and disk-spheroids grow only by a factor of $\sim1.5$. The latter is roughly consistent with the expected growth of disks in galaxy haloes, i.e  $R_e\propto H(z)^{-1}$ (black dashed-dotted lines in fig.~\ref{fig:med_props}) which comes from the theoretical assumption that disks are formed with a fixed fraction of mass and angular momentum of the parent halo (e.g. \citealp{1998MNRAS.295..319M,2004ApJ...600L.107F}). Spheroids grow at a faster rate as already pointed out by \cite{2013MNRAS.428.1715H} with a different selection, suggesting that some other mechanism takes place in these systems. \\

Sersic indices also do increase on average from $n\sim1.5$ to $n\sim4$, but again with different behaviors depending on morphology. The spheroids have $n>3$ and they increase up to $n\sim5$, confirming their bulge dominated morphologies at all epochs. On the other side, irregular disks have very low n values ($n\sim1$) while disk and disk+spheroids have rather constant intermediate values with ($n\sim1.5$ for disks and $n\sim2.5-3$ for disk+sph). This also confirms the validity of our morphological classification. Considering all these trends, the average observed increase of the Sersic index (black line in figure~\ref{fig:med_props}) is again better explained as a combination of morphological transformations from clumpy disks to regular systems which produces a growth of the bulge and an increase of the Sersic index together with the \emph{individual} increase of spheroids. The general increase of the Sersic index is also observed by \cite{2015arXiv150102800S} and might help to explain part of the evolution in the lensing profile of early-type galaxies. 

Axis ratios show little evolution with redshift but the absolute values change significantly with the morphological type. Spheroids have $b/a$ values close to 0.8 while all the other morphological classes present values of $0.5-0.6$, which again suggests that there are two families of objects following different evolutions. The measured values are also in good agreement with measurements in the local universe for similar morphologies \citep{2013MNRAS.436..697B}.

\begin{figure*}
\begin{center}
$\begin{array}{c c}
\includegraphics[width=0.40\textwidth]{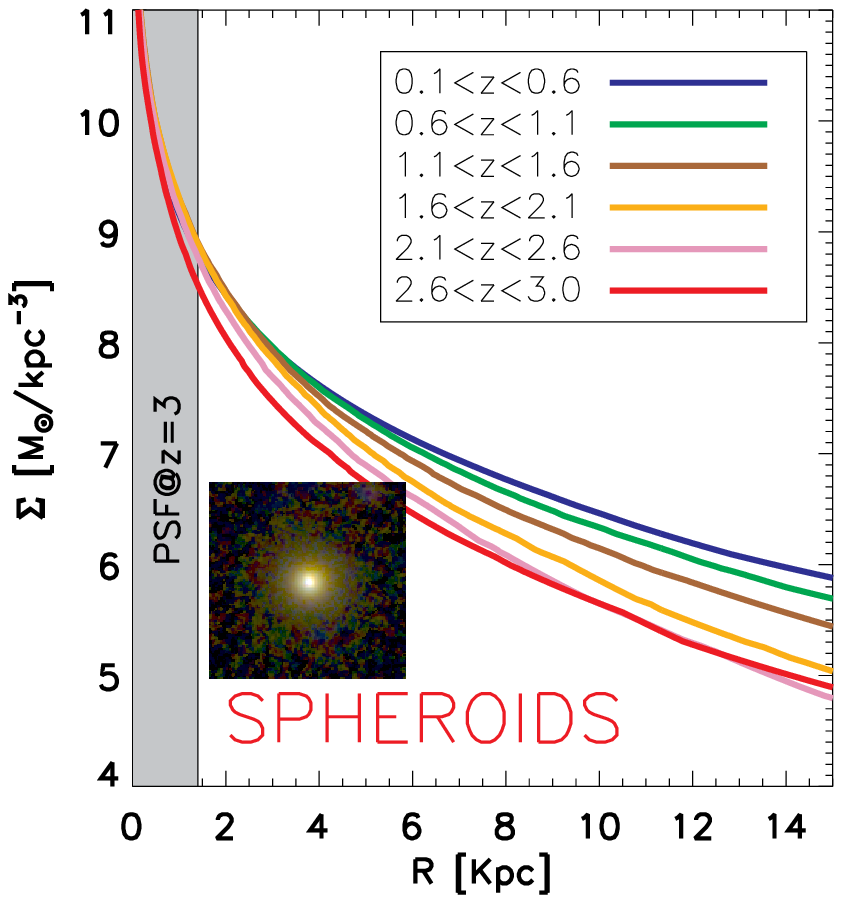} & \includegraphics[width=0.40\textwidth]{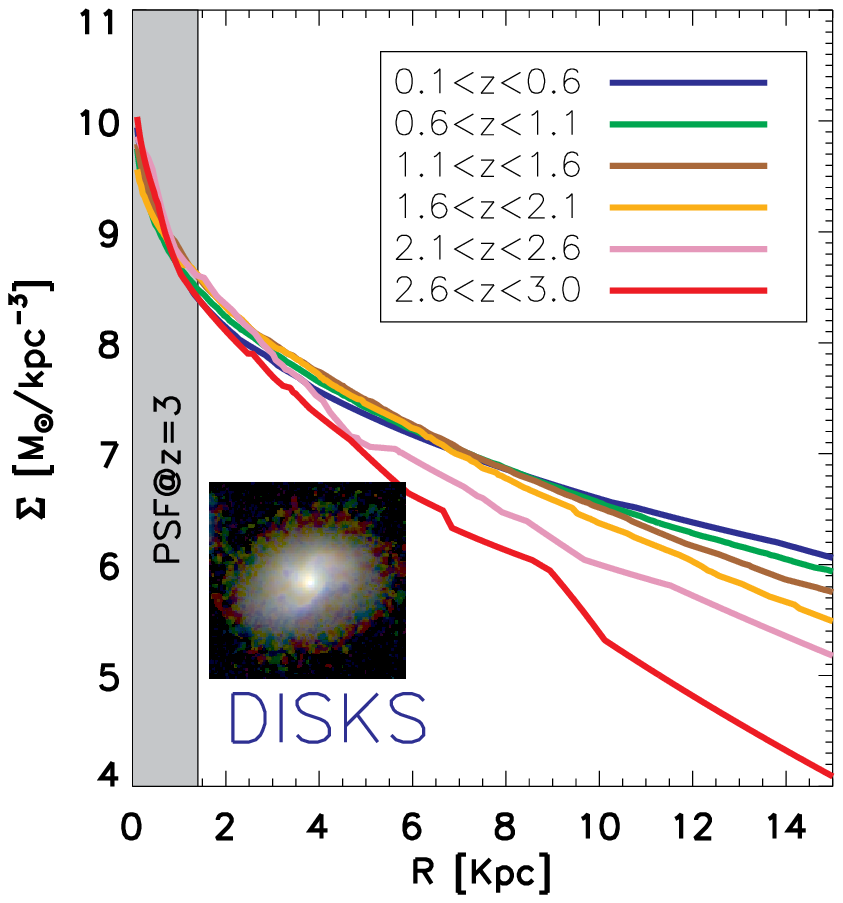}\\
\includegraphics[width=0.40\textwidth]{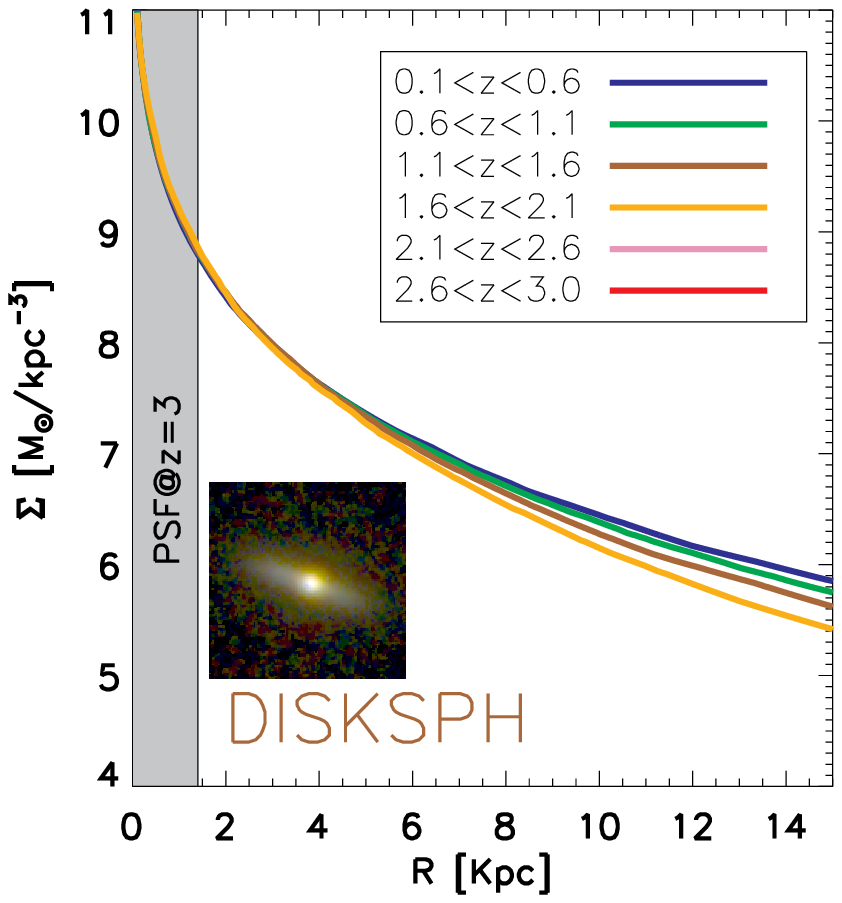} & \includegraphics[width=0.40\textwidth]{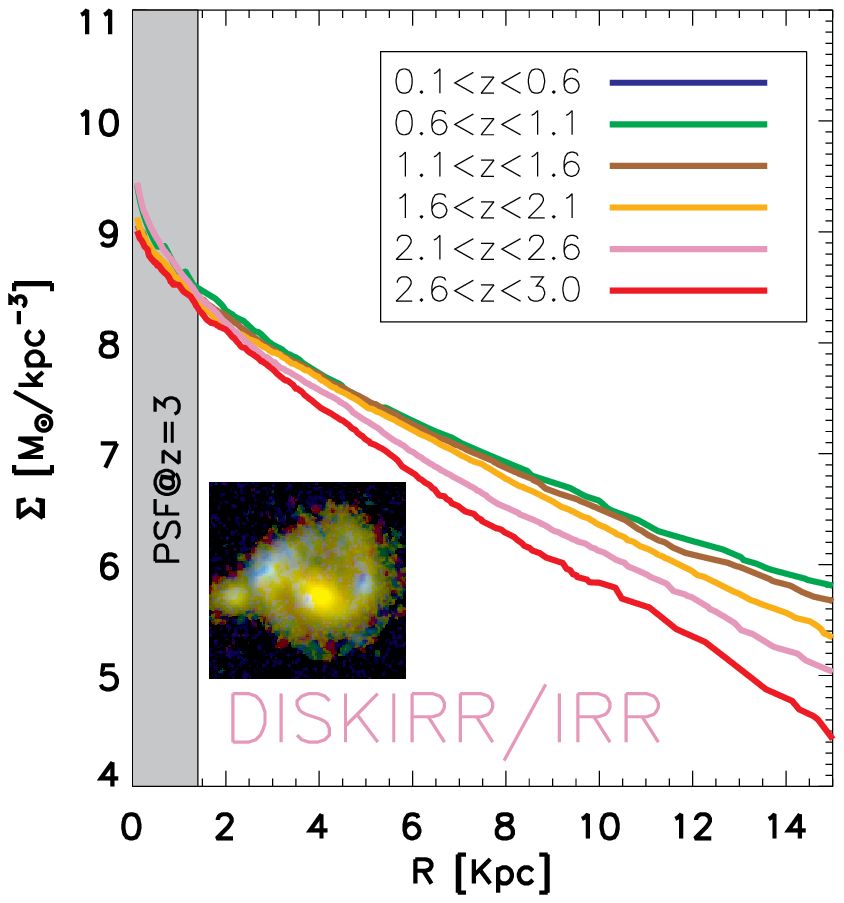}\\
\includegraphics[width=0.40\textwidth]{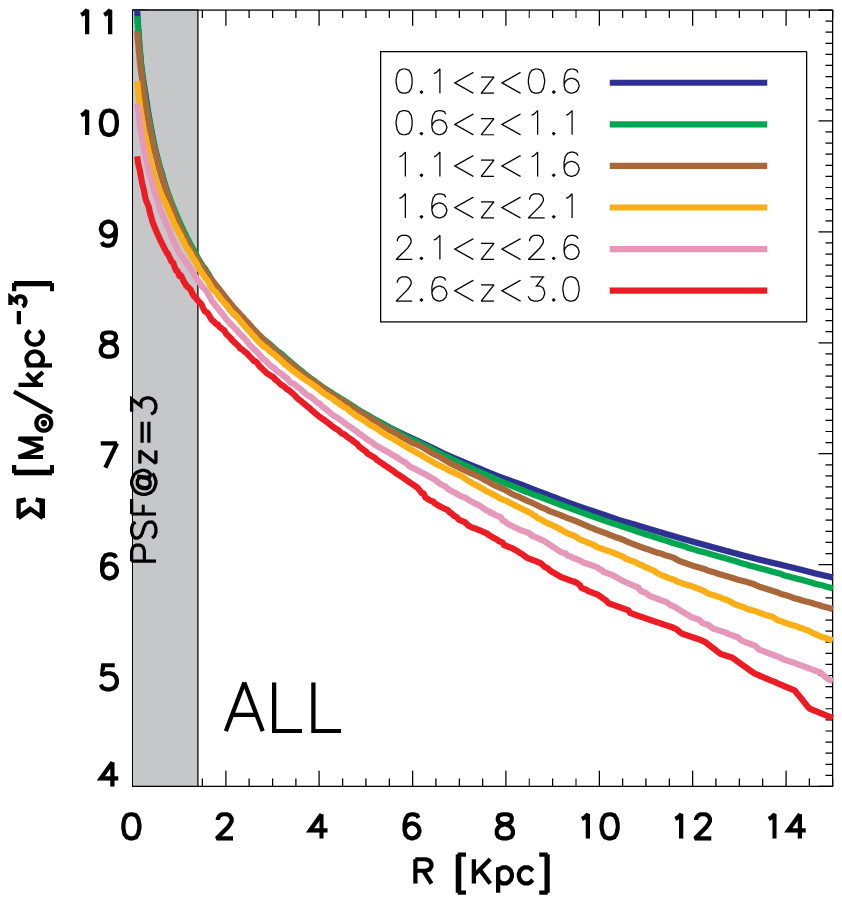} & \\
\end{array}$
\caption{Median stellar mass density profiles for different morphologies in different redshift bins as labelled. The top left panel, top right panel, middle left panel and middle right panel show the profiles for spheroids, disks, disk+spheroids and irregular disks respectively. The bottom panel shows the average profiles for all galaxies.} 
\label{fig:mass_profiles}
\end{center}
\end{figure*}

In figure~\ref{fig:mass_profiles} we analyze the total mass density profiles for different morphologies. As also done in \cite{2013ApJ...766...15P}, we compute the median mass density profile using the best fit Sersic models at different redshifts and convert them to stellar masses by normalizing by the stellar mass of each galaxy (see~\citealp{2013MNRAS.428..109S} for details). This procedure is clearly a first order approximation since it neglects any gradient in the stellar populations which could definitely modify the shape of the profiles (specially for the star forming galaxies). The figure clearly shows that spheroids are rapidly increasing their size at a faster rate than the average (bottom panel) with most of the action happening towards the galaxy outskirts through the addition of material. The central parts remain unaltered from $z\sim3$ (changes in the inner $1$ Kpc would not be detected given the PSF size). The evolution of the mass density profile for disks and disk+spheroids is less dramatic, resulting in a milder increase of their size, but the changes happen also at radii larger than  $3-4$ kpc. Interestingly, the clumpy-disks do show a significant evolution of their profile which could be interpreted as gas accretion happening in these objects and bringing material to the outskirts. These trends should however be taken with caution, specially for the star forming population, since firstly we neglect any stellar population gradient by construction when building the stacked mass profiles and secondly the profiles are  obtained through single component fits which might not be well adapted to reproduce the irregular light distribution of clumpy galaxies.

\section{Discussion: two channels of bulge growth}
\label{sec:disc}
The results presented in the previous sections seem to point out two different evolutionary tracks for massive galaxies ($log(M_*/M_\odot)\sim11.2\pm0.15$) and for the growth of their bulge component which are summarized in figure~\ref{fig:summary_plot}. As a matter of fact, the detailed analysis of the morphological properties of the progenitors from $z\sim3$ shows that there are 2 distinct families of galaxies with also different physical properties.  \\

\subsection{The \emph{nugget} track - Fast assembly}
 About $\sim30\%$ of massive galaxies had a spheroid morphology at $z\sim2.5$ - with no disk component- and this fraction does not evolve down to $z\sim0$. The quiescent fraction for the spheroid population is also rather high ($\sim50\%$) and their gas fraction low ($\sim10-15\%$) from $z\sim2$ and increases between $z\sim2-3$ which suggests that these galaxies are rapidly assembling at $z\sim2.5$ and above. The fraction of passive spheroids rises to almost $100\%$ at $z<0.5$ indicating that they are in the process of quenching in the epoch probed by this work but without significant alterations of their morphologies, in agreement with the findings of \cite{2013ApJ...765..104B,2014ApJ...791...52B} and \cite{2014arXiv1403.7524M} for dense regions. Their SFRs remain indeed well below the main-sequence of star formation at all epochs (fig.~\ref{fig:summary_plot}). The fast quenching is also accompanied by a rapid growth of their effective radii by a factor of $\sim5$, compared to a factor of $2$ growth in stellar mass, with most of the action happening in the galaxy outskirts ($R>4Kpc$). The Sersic index also increases from $n\sim3$ to $n\sim5$, even though it remains rather high at all epochs, confirming their bulge dominated morphology. They are therefore very similar to the dense-core galaxies identified by~\cite{2014ApJ...791...45V}. The increase of the Sersic index, is however not coupled to the gas content, at variance with what is observed for the average population~\citep{2014arXiv1412.3806P} which suggests an external driver. The fact that the number density and the morphologies do not change supports the idea that the reported growth is indeed an individual growth of these objects and that it is not driven by the morphological transformations or quenching of \emph{new} galaxies (progenitor bias). Otherwise, we would expect an increase of their abundance, since it is very unlikely that these galaxies will transform into another morphological class (although this cannot be fully excluded as discussed in section~\ref{sec:mergers}). An independent test for this statement would be a detailed analysis of the stellar population ages for this particular population which will be addressed in forthcoming work with higher resolution SEDs. Recall, that the size growth is even larger than the one measured for the overall population (i.e. factor of 2-3) which is in fact a convolution of different mechanisms as described below. The properties of these galaxies are therefore consistent with a formation of  the bulk of their stellar populations at high redshifts through violent disk instabilities or gas rich mergers (at $z\sim2.5$ their gas fraction is high and also their sSFR) which created their dense cores \citep{2014ApJ...791...52B,2014MNRAS.438.1870D} and a subsequent growth by the addition of material in the outskirts through, possibly, minor mergers. Figure~\ref{fig:summary_plot} summarizes the inferred evolution of these objects in the mass-size and $M_*$-SFR planes from $z\sim3$. It is worth emphasizing, that the size growth for these objects at later epochs ($z<1$) is still very pronounced, i.e. a factor of $\sim2$ with a minimum stellar mass growth. The global trend is nevertheless still compatible, at first order, with a minor merger driven growth as predicted by numerical and semi-empirical models (e.g. \citealp{2015arXiv150102800S} for slightly larger stellar masses). A more detailed comparison with the models predictions is however required at this stage.
 

\subsection{The clumpy track - Slow assembly}
\begin{figure*}
\begin{center}
$\begin{array}{c c}
\includegraphics[width=0.50\textwidth]{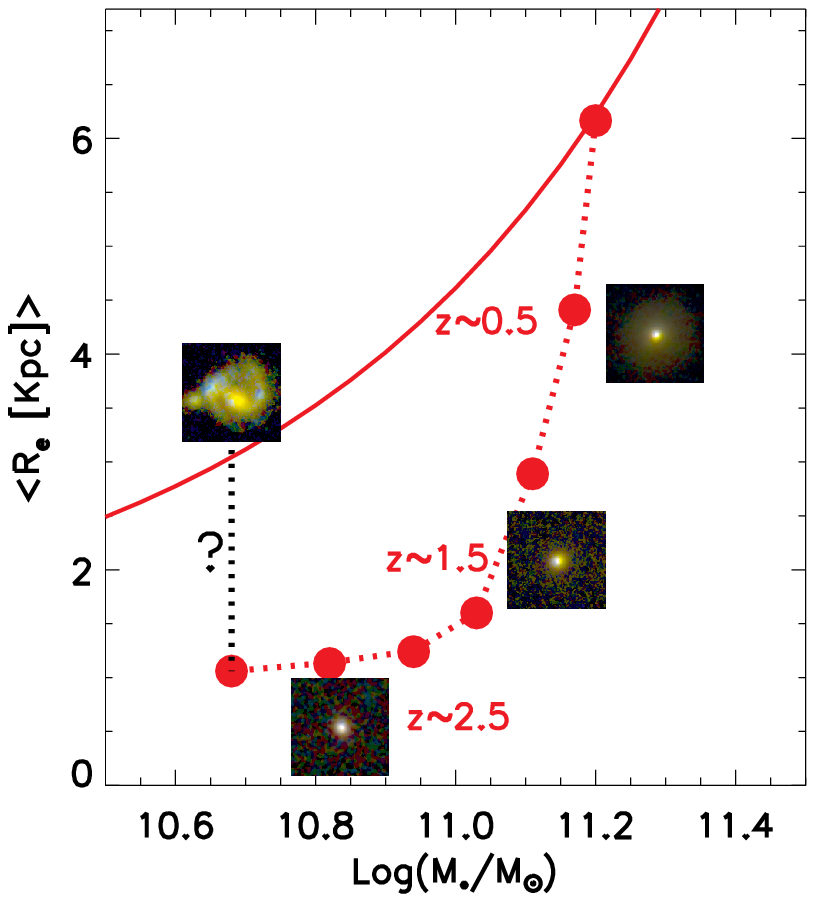} & \includegraphics[width=0.50\textwidth]{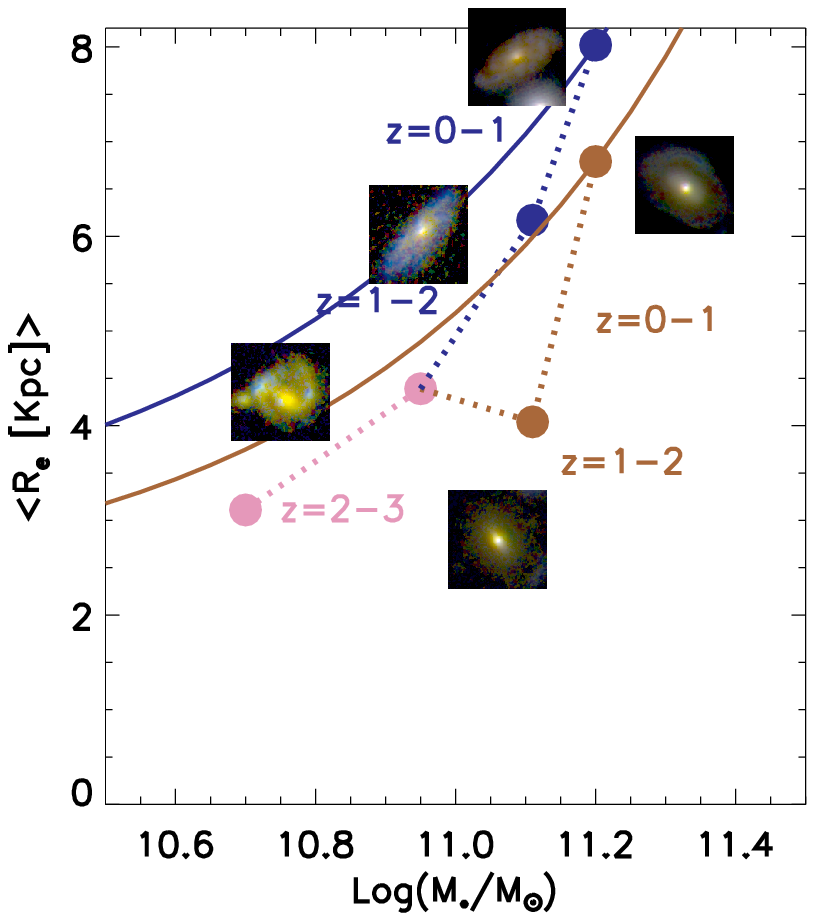}\\
\includegraphics[width=0.50\textwidth]{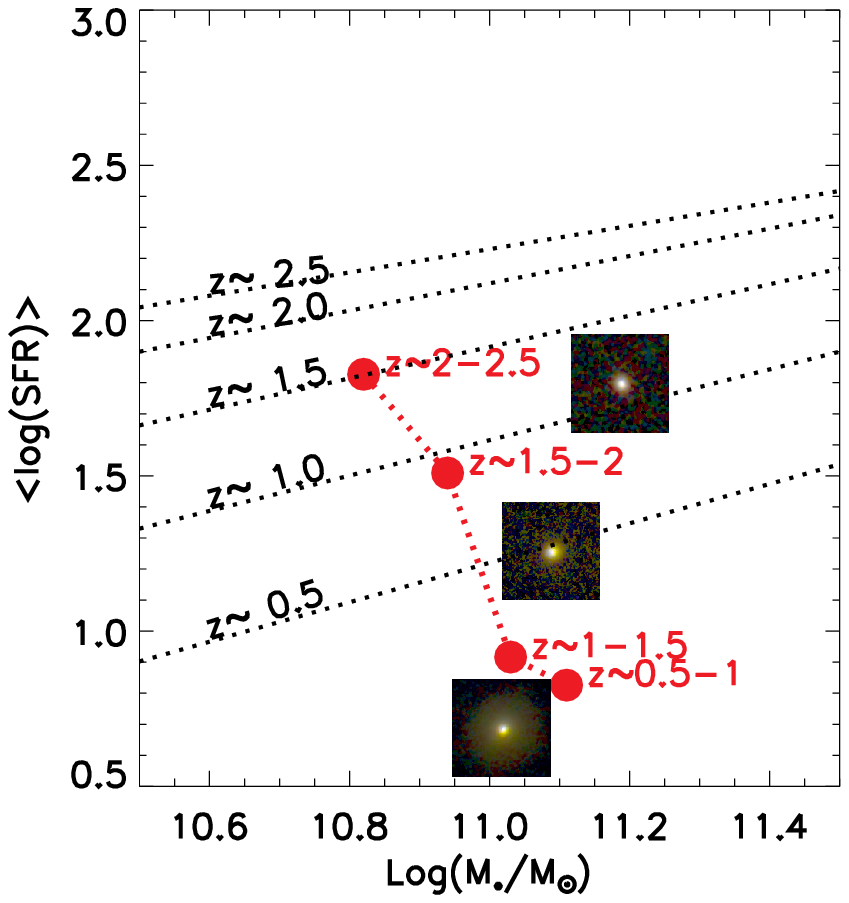} & \includegraphics[width=0.50\textwidth]{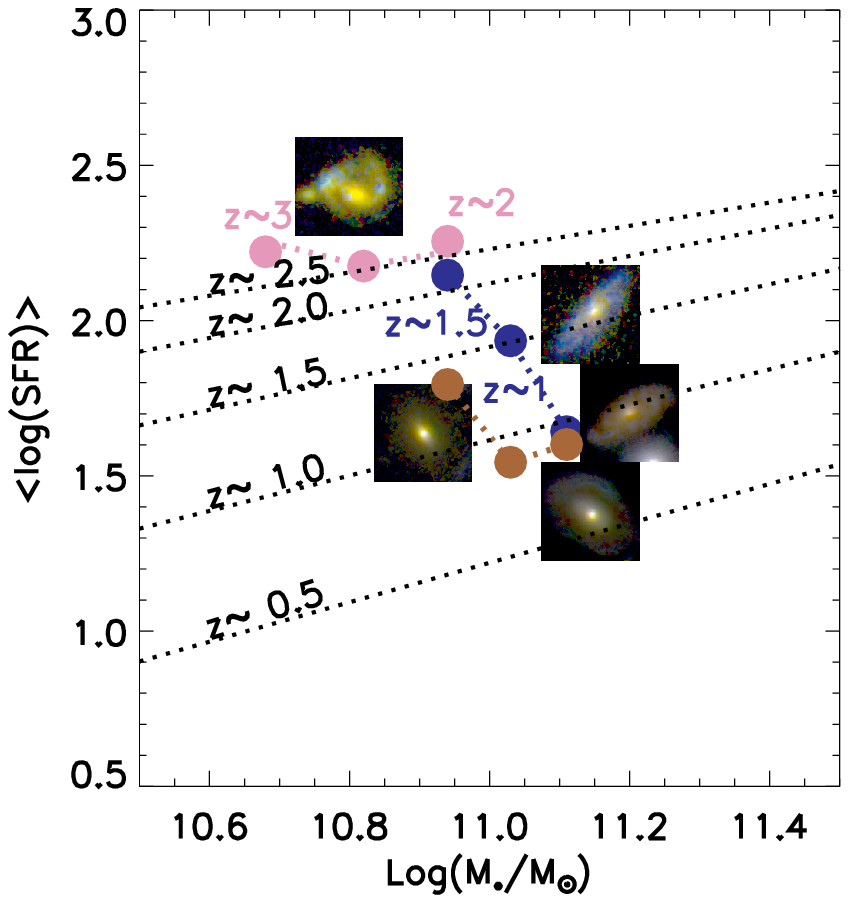}\\
\end{array}$
\caption{Expected evolution in the mass-size (top) and $M_*$-SFR (bottom) planes of the two channels of bulge growth (see text for details). The left panels show the evolution of spheroids. The right panels show the evolutionary track of clumpy disks. The red, yellow and blue solid lines in the top panels show the $z\sim0$ median mass-size relation from~\protect\cite{2014MNRAS.443..874B} for ellipticals, early-spirals and late-spirals respectively. The dashed lines in the bottom panels indicate the median star-forming main sequence at different redshifts from~\protect\cite{2012ApJ...754L..29W}} 
\label{fig:summary_plot}
\end{center}
\end{figure*}

The remaining $\sim60\%$ of the population is made of irregular/clumpy disks at $z\sim3$ which experience a rapid morphological transformation between $z=3$ and $z=1$ to give birth to very massive disks with a small bulge ($\sim20\%$) and to $40\%$ of galaxies with both a prominent bulge and a disk component. The evolution of the effective radii is more moderate than for spheroids and scales roughly with $H(z)^{-1}$, the expected growth of disks in DM haloes. Figure~\ref{fig:summary_plot} summarizes the inferred evolution of these objects in the mass-size and $M_*$-SFR planes from $z\sim3$. The transition between the clumpy-irregular morphologies to more Hubble sequence like galaxies happens mostly at $z>1$. Clumpy-irregular disks are characterized by high SFRs ($>100M_\odot.yr^{-1}$, slightly above the main-sequence at that epoch), high gas-fractions ($\sim60\%$) and low Sersic indices ($n\sim1$). Some of these objects ($\sim1/3$) will experience a smooth transition to become massive spirals with low B/T fractions. As a matter of fact, the properties of both families are very similar, in terms of gas fractions, Sersic index, SFRs and also effective radii. The other $\sim2/3$ will build a more prominent bulge ($n>2.5$) which roughly correspond to B/T of 50\%-75\% \citep{2012MNRAS.427.1666B}. The building-up of a larger bulge component results in a decrease of the effective radius following the concentration of mass towards the central regions. The emergence of the bulge is also tightly correlated with the decrease of the star formation activity and the decrease of the gas fractions which go from $50\%$ to $\sim10\%$, which make them depart from the star-formation main sequence (fig.~\ref{fig:summary_plot}). This evolution is consistent with the predictions of several numerical simulations (e.g. \citealp{2015arXiv150307660B,2009ApJ...707..250M}) which show how the growth of a bulge through clump migration is followed by a decrease of the star-formation activity (\emph{morphological quenching}) although the effect of feedback from a SMBH in the growing bulge could also produce similar effects. As a matter of fact, AGN feedback is known to contribute to the quenching of star formation (e.g., \citealp{1998A&A...331L...1S,2004ApJ...600..580G}) and it is also known to correlate with the mass of the bulge (e.g., \citealp{2013ARA&A..51..511K}) and therefore could also help explaining the quenching of bulge+disk systems which seems to be associated with the growth of the bulge. 

\subsection{Major mergers}
\label{sec:mergers}

This 2 track-scenario is obviously not the only possible explanation for the trends we observe. Namely, the constant number of spheroids could also be a result of clumpy gas rich galaxies being transformed into spheroids following a major merger event and spheroids regrowing a disk at the same rate (e.g. \citealp{2009A&A...507.1313H,2009ApJ...691.1168H}). Since the inferred gas fractions of irregular systems is high ($\sim40\%$), this is a plausible option. The scenario requires however a fine-tuning to keep these two effects (formation of spheroids and disk regrowth) at the same rate and also a high (major) merger fraction to keep producing spheroids. There have been several measurements of the major merger rate (1:4) of massive galaxies ($M_*/M_\odot\sim10^{11}$) in the recent literature. From the observational point of view, \cite{2011ApJ...742..103L} measure 1.6 mergers/galaxy between $z\sim0-3$ (extrapolating the quoted redshift evolution). This is in rather good agreement with \cite{2012ApJ...744...85M} who find 1.1 mergers/galaxy in the same period and also with \cite{2012A&A...548A...7L}. \cite{2012ApJ...747...34B,2009MNRAS.394L..51B} find a larger fraction (1.7 mergers/galaxy only between $z=1.7$ and $z=3$). Abundance matching based measurements (e.g. \citealp{2010ApJ...724..915H}) also find similar numbers (1.7 mergers/galaxy between $z=0$ and $z=3$), just as SAMs (e.g.~\citealp{2014MNRAS.444.1125C}) and numerical simulations (e.g.~\citealp{2014arXiv1411.2595K}). In this work we use the \cite{2013ApJ...777L..10B} model, which predicts $1.2$ major mergers/galaxy along the mass growth track shown in figure~\ref{fig:tracks}.

Considering these different measurements, it is certainly safe to assume that, on average, each galaxy in our sample experiences a major merger event in the redshift range explored. Assuming then that each merger event is enough to change the morphology, it is indeed possible to explain the number density decrease of irregular disks by mergers followed by disk rebuilding. We notice however that the \cite{2009ApJ...691.1168H} simulations focusing on disk rebuilding, predict that only equal mass mergers are able to create bulge dominated systems. Lower mass ration mergers (1:2-1:4) tend to create disk dominated systems which continue forming stars. In that respect, if mergers+disk rebuilding is the dominant channel, we would expect at all epochs an increasing fraction of star-forming regular disks and a minor fraction of quiescent bulge-dominated systems. The opposite is actually observed.

In addition to this, we do observe that the median ages of the spheroids in our sample estimated through SED fitting, monolithically increase from $\sim$0.5 Gyrs at $z>2$ to $\sim$2.5 Gyr at $z\sim0.5$ (roughly consistent with the time between these 2 redshift bins). On the other hand, the ages of disks dominated systems tend to stay rather young ($<1 Gyr$) at all epochs due to the sustained star formation. If the dominant process to create the $30\%$ spheroids we observe is merging, we would not expect a strong increase of the ages of spheroids which we actually seem to observe. Given the known degeneracies affecting age determination from broad band photometry, these trends need to be taken with caution. However they point towards an early formation of the spheroid population. A further check of the proposed bulge growth tracks would therefore imply accurate age estimations of the bulge components in the different galaxy types. This requires bulge-to-disk decompositions and high resolution SED fitting of the different components which is on-going.

\section{Summary}
\label{sec:summ}
We have analyzed the morphological, structural and star-formation properties of the progenitors of massive galaxies ($M_*/M_\odot\sim10^{11.2\pm0.3}$) from $z\sim3$. The progenitors are selected using abundance matching to take into account the expected mass growth in the redshift range probed in this work. The latter selection is a key point of the present work and is clearly subjected to important assumptions (i.e. halo mass functions, galaxy stellar mass functions and halo occupation distribution functions) as explained in the previous sections. It is worth emphasizing however that selecting galaxies at fixed stellar mass (i.e. assuming the extreme scenario in which galaxies do not grow in mass) results in very similar trends that the ones reported here. \\

The main new ingredient of this work is the addition of accurate visual-like morphologies which helps better understanding the different evolutionary tracks leading to the present day Hubble sequence. We have defined 5 main morphological types which quantify the presence or not of a bulge/disk component and the presence or absence of irregularities in the light profile. We then have explored, the abundances, star formation rates, quiescent fractions, gas fractions and structural properties for each morphological type. \\

Our main results are the following:

\begin{itemize}

\item The morphologies of massive galaxies significantly change from $z\sim3$. At $z<1$, these galaxies are made by $40\%$ of pure spheroids, $40\%$ bulge+disk galaxies (early spirals and lenticulars) and $20\%$ of massive disks. At $z\sim3$ there is still a $40\%$ of spheroids, but the remaining $60\%$ is made of irregular/clumpy disks or disturbed galaxies. Most of the morphological transformations take place at $z>1$.

\item As reported in previous works, the overall population of massive galaxies rapidly quenches from $z\sim3$ to $z\sim0$, i.e. the quiescent fraction increases from $\sim20\%$ to $\sim80\%$, the median SFR decreases from $\sim100M_\odot.yr^{-1}$ to $\sim25M_\odot.yr^{-1}$ and the gas fractions go from $\sim40\%$ to $\sim15\%$. When inspected at fixed visual morphology the trends are very different. The quiescent fraction in the spheroid population is already high at $z\sim3$, i.e. $60\%$ and increases to an almost $100\%$ value. The quiescent fraction for disks and disk irregulars remains low ($<20\%$) at all epochs while for bulge+disk objects the fraction appears to be constant too at a value of $40-50\%$. These trends suggest that the overall increase of the quenching fraction for the whole population can be explained by a combination of the quenching of the spheroid population with the morphological transformation from clumpy/irregular disk to early-spiral/S0.

\item When considering the overall population, without morphology distinction, we measure an increase of the average effective radius by a factor of $2-3$ as well as an increase of the Sersic index from $n\sim1.5$ to $n\sim4$, as reported in the recent literature for similar selections. The evolution of the average mass density profile is also in agreement with an inside-out growth. The evolution of the average size seems to have 2 different regimes, from $z\sim3$ to $z\sim1.5$, there is almost no significant change of the effective radius while the bulk of the growth happens from $z\sim1.5$ to $z\sim0$. At fixed morphologies, spheroids do grow by a factor of $5-6$ from $z\sim3$ and increase their Sersic index from $n\sim3$ to $n\sim5$. On the other hand, irregular disks and disks grow by a factor of $\sim1.5$ and keep a rather constant Sersic index ($n<2$ for disks and disk+irr) and $n\sim2.5-3$ for disk+bulge galaxies. The two different phases in the average growth are better explained if morphological transformations are taken into account. In the first phase, there is a rapid morphological transformation from clumpy disks to bulge+disk galaxies which results in a slight decrease of the effective radius as a consequence of the mass concentration towards the inner regions of the galaxy. Even though, spheroids and clumpy disks increase their size in this period, the rapid decrease of the number density of the latter seems to compensate this growth and results in no evolution of the average size. During the second phase, from $z\sim1.5$, the morphological mixing remains roughly constant, but the size growth increases by a factor of $\sim2-3$ on average. This growth is therefore better explained by the individual growth of disk/disk+bulge galaxies which grow by a factor of $\sim1.5$ and the growth of the spheroids which increase their effective radius by a factor of $\sim4$.   

\end{itemize}

The above results suggest to different channels for the bulge growth in the massive end of the Hubble sequence:

\begin{enumerate}

\item A \emph{nugget} track (fast assembly) followed by $30-40\%$ of the population of massive galaxies. Galaxies formed that way, formed the bulk of the stars at $z>2.5$ and also acquired their spheroidal morphology at these early epochs possible through violent disk instabilities (and/or mergers) which rapidly bring gas into the central parts. At $z<2$, they have already low gas fractions, low SFRs, high Sersic indices and $\sim60\%$ of them are classified as quiescent. They are however very compact with median effective radii of $\sim5$ kpc. Between $z\sim3$ and $z\sim0.5$ they practically completely stop forming stars while they increase their size by a factor of $\sim5$ and their Sersic index from $n\sim3$ to $n\sim5$ keeping their global visual aspect unaffected. The growth is decoupled of the gas content and the SFRs which remain low at all epochs and mostly happens in the galaxy outskirts, suggesting an ex-situ driven growth.

\item A \emph{clumpy} track (slow assembly), followed by $60-70\%$ of the population of massive galaxies at $z\sim0$. These galaxies were clumpy/irregular star-forming disks ($SFR>100M_\odot.yr^{-1}$) at $z\sim2-3$. From $z\sim3$ to $z\sim1$ they experience a rapid morphological transformation leading to relaxed systems (at least in terms of their visual aspect) and to the emergence of a bugle component of variable size ($\sim2/3$ seem to develop a large bulge component while the remaining $1/3$ keep a disk dominated morphology). The morphological transformation is accompanied by a decrease of the SFR (reaching $\sim50M_\odot.yr^{-1}$) and the gas fraction (going down to $\sim15\%$), more dramatic for galaxies developing a larger bugle as well as by an increase of the Sersic index (from $n\sim1$ to $n\sim2.5-3$) and a decrease of the effective radius because of the mass being concentrated towards the inner regions. This is in good agreement with the predictions of numerical simulations in which the bulge component is built from the migration of clumps and the stabilization of the disk results in a decrease of the star formation rate (\emph{morphological quenching}) although the possible effect of a SMBH should also be considered. Major merger events followed by a disk rebuilding event could also contribute to transform irregular systems but it is unlikely to be the dominant channel in the mass range explored in this paper. Below $z\sim1$, the well known massive end of the Hubble sequence is in place and the galaxy properties change only marginally. Their effective radii grow in fact at a rate roughly consistent with $H(z)^{-1}$ the expected growth due to the hierarchical assembly of haloes (e.g. \citealp{2014MNRAS.441.1570S}).

\end{enumerate}

{\bf Acknowledgments: } We would like to thank the referee for his useful and constructive report which clearly helped improving the manuscript. MHC really thanks P. Behroozi for kindly sharing the mass growth and merger rates of massive galaxies used in this paper. PGPG acknowledges funding from Spanish Government MINECO Grant AYA2012-31277. GCV gratefully acknowledges funding from CONICYT (Chile) through their Doctoral Scholarship. This work has made use of the Rainbow Cosmological Surveys Database, which is operated by the Universidad Complutense de Madrid (UCM), partnered with the University of California Observatories at Santa Cruz (UCO/Lick,UCSC). This work is based on observations taken by the 3D-HST Treasury Program (GO 12177 and 1232) with the NASA/ESA HST, which is operated by the Association of Universities for Research in Astronomy, Inc, under NASA contract NAS5-26555.

\clearpage

\end{document}